\documentclass[znote,zbstdefault]{./zeus_paper}

\chardef\usc=95
\chardef\til=126
\catcode`\@=11 
\DeclareRobustCommand\xdotspace{\futurelet\@let@token\@xdotspace}
\def\@xdotspace{%
  \ifx\@let@token.\else
  \ifx\@let@token\bgroup.\else
  \ifx\@let@token\egroup.\else
  \ifx\@let@token\/.\else
  \ifx\@let@token\ .\else
  \ifx\@let@token~.\else
  \ifx\@let@token!.\else
  \ifx\@let@token,.\else
  \ifx\@let@token:.\else
  \ifx\@let@token;.\else
  \ifx\@let@token?.\else
  \ifx\@let@token/.\else
  \ifx\@let@token'.\else
  \ifx\@let@token).\else
  \ifx\@let@token-.\else
  \ifx\@let@token\@xobeysp.\else
  \ifx\@let@token\space.\else
  \ifx\@let@token\@sptoken.\else
   .\space
   \fi\fi\fi\fi\fi\fi\fi\fi\fi\fi\fi\fi\fi\fi\fi\fi\fi\fi}
\catcode`\@=12 

\newcommand{\stru}[2]{%
   \relax\ifmmode\hbox{\vrule height#1 depth#2 width0pt}%
   \else\vrule height#1 depth#2 width0pt\fi}

\newcommand{\Ronum}[1]{\uppercase\expandafter{\romannumeral#1}}
\newcommand{\ronum}[1]{\expandafter{\romannumeral#1}}
\DeclareRobustCommand{\LaTeXZ}{%
  \LaTeX\kern-.05em4\kern-.1em
  {\raisebox{-0.2ex}{$\scriptstyle\text{ZEUS}$}}\xspace}

\DeclareMathAlphabet{\mathbf}{OT1}{cmr}{bx}{sl}
\newcommand{\eVdist}{\kern-0.06667em}

\newcommand{\gev}{{\,\text{Ge}\eVdist\text{V\/}}}

\newcommand{\Tesla}{\,\text{T}}

\newcommand{\slashfrac}[2]{%
  \raisebox{0.5ex}{\ensuremath #1}\kern-0.12em/\kern-0.08em
  \raisebox{-.8ex}{\ensuremath #2}}

\newcommand{\sqr}[3]{%
    {\vcenter{\hrule height.#3ex\hbox{\vrule width.#2ex height#1ex
     \kern#1ex\vrule width.#3ex}\hrule height.#2ex}}}

\catcode`\@=11 
\newcommand{\parenbar}{\mathpalette\p@renb@r}
\def\p@renb@r#1#2{\vbox{%
  \ifx#1\scriptscriptstyle \dimen@.7em\dimen@ii.2em\else
  \ifx#1\scriptstyle \dimen@.8em\dimen@ii.25em\else
  \dimen@1em\dimen@ii.4em\fi\fi \offinterlineskip
  \ialign{\hfill##\hfill\cr
    \vbox{\hrule width\dimen@ii}\cr
    \noalign{\vskip-.3ex}%
    \hbox to\dimen@{$\mathchar300\hfil\mathchar301$}\cr
    \noalign{\vskip-.3ex}%
    $#1#2$\cr}}}
\catcode`\@=12 

\newcommand{\IP}{{\rm I$\kern-0.01667em$P}\xspace}

\mathchardef\qsm=63
\mathchardef\pls=43
\mathchardef\mns=512
\mathchardef\plm=518
\mathchardef\eql=61
\mathchardef\smallleft=300
\mathchardef\smallright=301
\mathchardef\les=316
\mathchardef\gre=318
\mathchardef\leq=532
\mathchardef\grq=533

\catcode`\@=11 
\newcounter{pict@width}
\newcounter{pict@height}
\newlength{\pict@scale}
\newcommand{\psfigadd}[4]{%
\setcounter{pict@width}{1*\ratio{#2+\pict@scale/2}{\pict@scale}}
\setcounter{pict@height}{1*\ratio{#3+\pict@scale/2}{\pict@scale}}
\setlength{\unitlength}{\pict@scale}
\hbox to #2{\hspace{-\fill}\begin{picture}(\thepict@width,\thepict@height)
\put(0,0){\psfig{figure=#1,width=#2,height=#3,clip=}}
\SetScale{0.283466457}
\SetWidth{1.763889}
{#4}
\end{picture}}
}
\newcounter{pict@widthfst}
\newcounter{pict@widthscd}
\newcounter{pict@widthtot}
\newcommand{\psfigaddtwo}[7]{%
\setcounter{pict@widthfst}{1*\ratio{#2+\pict@scale/2}{\pict@scale}}
\setcounter{pict@widthscd}{1*\ratio{#2+#4+\pict@scale/2}{\pict@scale}}
\setcounter{pict@widthtot}{1*\ratio{#2+#4+#6+\pict@scale/2}{\pict@scale}}
\setcounter{pict@height}{1*\ratio{#3+\pict@scale/2}{\pict@scale}}
\setlength{\unitlength}{\pict@scale}
\hbox{\hspace{-\fill}\begin{picture}(\thepict@widthtot,\thepict@height)
\put(0,0){\psfig{figure=#1,width=#2,height=#3,clip=}}
\put(\thepict@widthscd,0){\psfig{figure=#5,width=#6,height=#3,clip=}}
\SetScale{0.283466457}
\SetWidth{1.763889}
{#7}
\end{picture}}
}
\newcommand{\psfigror}[4]{%
\setcounter{pict@width}{1*\ratio{#2+\pict@scale/2}{\pict@scale}}
\setcounter{pict@height}{1*\ratio{#3+\pict@scale/2}{\pict@scale}}
\setlength{\unitlength}{\pict@scale}
\hbox{\begin{picture}(\thepict@width,\thepict@height)
\put(0,\thepict@height){\psfig{figure=#1,width=#3,height=#2,clip=,angle=270}}
\SetScale{0.283466457}
\SetWidth{1.763889}
{#4}
\end{picture}}
}
\newcommand{\psfigrol}[4]{%
\setcounter{pict@width}{1*\ratio{#2+\pict@scale/2}{\pict@scale}}
\setcounter{pict@height}{1*\ratio{#3+\pict@scale/2}{\pict@scale}}
\setlength{\unitlength}{\pict@scale}
\hbox{\begin{picture}(\thepict@width,\thepict@height)
\put(0,0){\psfig{figure=#1,width=#3,height=#2,clip=,angle=90}}
\SetScale{0.283466457}
\SetWidth{1.763889}
{#4}
\end{picture}}
}
\catcode`\@=12 
\newlength\listtextwidth

\catcode`\@=11 
\newlength{\@tabfninsert}
\newlength{\@tabfnwidth}
\newcommand{\tabfootnote}[2]{%
  \setlength{\@tabfninsert}{0.8em}
  \setlength{\@tabfnwidth}{\textwidth}
  \addtolength{\@tabfnwidth}{-\@tabfninsert}
  \addtolength{\@tabfnwidth}{-0.4em}
  \noindent\makebox[\@tabfninsert][r]{\footnotesize$^{#1}$\hfil}\hfill%
  \parbox[t]{\@tabfnwidth}{\footnotesize #2\hfill}}
\catcode`\@=12 

\def\Cref#1{Chapter~\ref{#1}}                   
\def\cref#1{chapter~\ref{#1}}

\def\citeCTD{{\cite{%
nim:a279:290,*npps:b32:181,*nim:a338:254%
}}\xspace}
\def\citeMVD{{\cite{%
nim:a581:656%
}}\xspace}
\def\citeCAL{{\cite{%
nim:a309:77,*nim:a309:101,*nim:a321:356,*nim:a336:23%
}}\xspace}

\def\citePCAL{{\cite{%
desy-92-066,*zfp:c63:391,*acpp:b32:2025%
}}\xspace}

\usepackage[english]{babel}
\usepackage{longtable,lscape}
\usepackage{subfig}
\usepackage{fixltx2e}
\usepackage{changepage}

\begin{document}


\prepnum{DESY--14--053}

\title{
Deep inelastic cross-section measurements at large \boldmath{$y$} with the ZEUS detector at HERA
}                                                       
                    
\author{ZEUS Collaboration\\}

\abstract{
The reduced cross sections for $e^{+}p$ deep inelastic scattering have been 
measured with the ZEUS detector at HERA at three different centre-of-mass energies, 
$318$, $251$ and $225$ GeV.
The cross sections, measured double 
differentially in Bjorken $x$ and the virtuality, $Q^2$, were obtained 
in the region $0.13\ \leq\ y\ \leq\ 0.75$,
where $y$ denotes the inelasticity
and $5\ \leq\ Q^2\ \leq\ 110$~GeV$^2$.
The proton structure functions $F_2$ and $F_L$ were
extracted from the measured cross sections.  \\
\vspace{3cm}\\
\centerline{\emph{Dedicated to our friend and colleague Alexander Proskuryakov.}}
}
\makezeustitle

%
%
%
%

\def\3{\ss}
\pagenumbering{Roman}
                                                   %
\begin{center}
{                      \Large  The ZEUS Collaboration              }
\end{center}

{\small


        {\raggedright
H.~Abramowicz$^{27, u}$, 
I.~Abt$^{21}$, 
L.~Adamczyk$^{8}$, 
M.~Adamus$^{34}$, 
R.~Aggarwal$^{4, a}$, 
S.~Antonelli$^{2}$, 
O.~Arslan$^{3}$, 
V.~Aushev$^{16, 17, o}$, 
Y.~Aushev,$^{17, o, p}$, 
O.~Bachynska$^{10}$
% ...
        }

\newpage


\makebox[3em]{$^{1}$}
\begin{minipage}[t]{14cm}
{\it INFN Bologna, Bologna, Italy}~$^{A}$

\end{minipage}\\
\makebox[3em]{$^{2}$}
\begin{minipage}[t]{14cm}
{\it University and INFN Bologna, Bologna, Italy}~$^{A}$

\end{minipage}\\
\makebox[3em]{$^{3}$}
\begin{minipage}[t]{14cm}
{\it Physikalisches Institut der Universit\"at Bonn,
Bonn, Germany}~$^{B}$

\end{minipage}\\
\makebox[3em]{$^{4}$}
\begin{minipage}[t]{14cm}
{\it Panjab University, Department of Physics, Chandigarh, India}

\end{minipage}\\
\makebox[3em]{$^{5}$}
\begin{minipage}[t]{14cm}
{\it Calabria University,
Physics Department and INFN, Cosenza, Italy}~$^{A}$

\end{minipage}\\
\makebox[3em]{$^{6}$}
\begin{minipage}[t]{14cm}
{\it National Centre for Particle Physics, Universiti Malaya, 50603 Kuala Lumpur, Malaysia}~$^{C}$

\end{minipage}\\
\makebox[3em]{$^{7}$}
\begin{minipage}[t]{14cm}
{\it The Henryk Niewodniczanski Institute of Nuclear Physics, Polish Academy of \\
Sciences, Krakow, Poland}~$^{D}$

\end{minipage}\\
\makebox[3em]{$^{8}$}
\begin{minipage}[t]{14cm}
{\it AGH-University of Science and Technology, Faculty of Physics and Applied Computer
Science, Krakow, Poland}~$^{D}$

\end{minipage}\\
\makebox[3em]{$^{9}$}
\begin{minipage}[t]{14cm}
{\it Department of Physics, Jagellonian University, Cracow, Poland}

\end{minipage}\\
\makebox[3em]{$^{10}$}
\begin{minipage}[t]{14cm}
{\it Deutsches Elektronen-Synchrotron DESY, Hamburg, Germany}

\end{minipage}\\
\makebox[3em]{$^{11}$}
\begin{minipage}[t]{14cm}
{\it Deutsches Elektronen-Synchrotron DESY, Zeuthen, Germany}

\end{minipage}\\
\makebox[3em]{$^{12}$}
\begin{minipage}[t]{14cm}
{\it School of Physics and Astronomy, University of Glasgow,
Glasgow, United Kingdom}~$^{E}$

\end{minipage}\\
\makebox[3em]{$^{13}$}
\begin{minipage}[t]{14cm}
{\it Hamburg University, Institute of Experimental Physics, Hamburg,
Germany}~$^{F}$

\end{minipage}\\
\makebox[3em]{$^{14}$}
\begin{minipage}[t]{14cm}
{\it Institute of Particle and Nuclear Studies, KEK,
Tsukuba, Japan}~$^{G}$

\end{minipage}\\
\makebox[3em]{$^{15}$}
\begin{minipage}[t]{14cm}
{\it Institute of Physics and Technology of Ministry of Education and
Science of Kazakhstan, Almaty, Kazakhstan}

\end{minipage}\\
\makebox[3em]{$^{16}$}
\begin{minipage}[t]{14cm}
{\it Institute for Nuclear Research, National Academy of Sciences, Kyiv, Ukraine}

\end{minipage}\\
\makebox[3em]{$^{17}$}
\begin{minipage}[t]{14cm}
{\it Department of Nuclear Physics, National Taras Shevchenko University of Kyiv, Kyiv, Ukraine}

\end{minipage}\\
\makebox[3em]{$^{18}$}
\begin{minipage}[t]{14cm}
{\it Department of Physics, McGill University,
Montr\'eal, Qu\'ebec, Canada H3A 2T8}~$^{H}$

\end{minipage}\\
\makebox[3em]{$^{19}$}
\begin{minipage}[t]{14cm}
{\it Meiji Gakuin University, Faculty of General Education,
Yokohama, Japan}~$^{G}$

\end{minipage}\\
\makebox[3em]{$^{20}$}
\begin{minipage}[t]{14cm}
{\it Lomonosov Moscow State University, Skobeltsyn Institute of Nuclear Physics,
Moscow, Russia}~$^{I}$

\end{minipage}\\
\makebox[3em]{$^{21}$}
\begin{minipage}[t]{14cm}
{\it Max-Planck-Institut f\"ur Physik, M\"unchen, Germany}

\end{minipage}\\
\makebox[3em]{$^{22}$}
\begin{minipage}[t]{14cm}
{\it Department of Physics, University of Oxford,
Oxford, United Kingdom}~$^{E}$

\end{minipage}\\
\makebox[3em]{$^{23}$}
\begin{minipage}[t]{14cm}
{\it INFN Padova, Padova, Italy}~$^{A}$

\end{minipage}\\
\makebox[3em]{$^{24}$}
\begin{minipage}[t]{14cm}
{\it Dipartimento di Fisica dell' Universit\`a and INFN,
Padova, Italy}~$^{A}$

\end{minipage}\\
\makebox[3em]{$^{25}$}
\begin{minipage}[t]{14cm}
{\it Polytechnic University, Tokyo, Japan}~$^{G}$

\end{minipage}\\
\makebox[3em]{$^{26}$}
\begin{minipage}[t]{14cm}
{\it Dipartimento di Fisica, Universit\`a `La Sapienza' and INFN,
Rome, Italy}~$^{A}$

\end{minipage}\\
\makebox[3em]{$^{27}$}
\begin{minipage}[t]{14cm}
{\it Raymond and Beverly Sackler Faculty of Exact Sciences, School of Physics, \\
Tel Aviv University, Tel Aviv, Israel}~$^{J}$

\end{minipage}\\
\makebox[3em]{$^{28}$}
\begin{minipage}[t]{14cm}
{\it Department of Physics, Tokyo Institute of Technology,
Tokyo, Japan}~$^{G}$

\end{minipage}\\
\makebox[3em]{$^{29}$}
\begin{minipage}[t]{14cm}
{\it Universit\`a di Torino and INFN, Torino, Italy}~$^{A}$

\end{minipage}\\
\makebox[3em]{$^{30}$}
\begin{minipage}[t]{14cm}
{\it Universit\`a del Piemonte Orientale, Novara, and INFN, Torino,
Italy}~$^{A}$

\end{minipage}\\
\makebox[3em]{$^{31}$}
\begin{minipage}[t]{14cm}
{\it Department of Physics, University of Toronto, Toronto, Ontario,
Canada M5S 1A7}~$^{H}$

\end{minipage}\\
\makebox[3em]{$^{32}$}
\begin{minipage}[t]{14cm}
{\it Physics and Astronomy Department, University College London,
London, United Kingdom}~$^{E}$

\end{minipage}\\
\makebox[3em]{$^{33}$}
\begin{minipage}[t]{14cm}
{\it Faculty of Physics, University of Warsaw, Warsaw, Poland}

\end{minipage}\\
\makebox[3em]{$^{34}$}
\begin{minipage}[t]{14cm}
{\it National Centre for Nuclear Research, Warsaw, Poland}

\end{minipage}\\
\makebox[3em]{$^{35}$}
\begin{minipage}[t]{14cm}
{\it Department of Particle Physics and Astrophysics, Weizmann
Institute, Rehovot, Israel}

\end{minipage}\\
\makebox[3em]{$^{36}$}
\begin{minipage}[t]{14cm}
{\it Department of Physics, York University, Ontario, Canada M3J 1P3}~$^{H}$

\end{minipage}\\
\vspace{30em} \pagebreak[4]


\makebox[3ex]{$^{ A}$}
\begin{minipage}[t]{14cm}
 supported by the Italian National Institute for Nuclear Physics (INFN) \
\end{minipage}\\
\makebox[3ex]{$^{ B}$}
\begin{minipage}[t]{14cm}
 supported by the German Federal Ministry for Education and Research (BMBF), under
 contract No. 05 H09PDF\
\end{minipage}\\
\makebox[3ex]{$^{ C}$}
\begin{minipage}[t]{14cm}
 supported by HIR grant UM.C/625/1/HIR/149 and UMRG grants RU006-2013, RP012A-13AFR and RP012B-13AFR from
 Universiti Malaya, and ERGS grant ER004-2012A from the Ministry of Education, Malaysia\
\end{minipage}\\
\makebox[3ex]{$^{ D}$}
\begin{minipage}[t]{14cm}
 supported by the National Science Centre under contract No. DEC-2012/06/M/ST2/00428\
\end{minipage}\\
\makebox[3ex]{$^{ E}$}
\begin{minipage}[t]{14cm}
 supported by the Science and Technology Facilities Council, UK\
\end{minipage}\\
\makebox[3ex]{$^{ F}$}
\begin{minipage}[t]{14cm}
 supported by the German Federal Ministry for Education and Research (BMBF), under
 contract No. 05h09GUF, and the SFB 676 of the Deutsche Forschungsgemeinschaft (DFG) \
\end{minipage}\\
\makebox[3ex]{$^{ G}$}
\begin{minipage}[t]{14cm}
 supported by the Japanese Ministry of Education, Culture, Sports, Science and Technology
 (MEXT) and its grants for Scientific Research\
\end{minipage}\\
\makebox[3ex]{$^{ H}$}
\begin{minipage}[t]{14cm}
 supported by the Natural Sciences and Engineering Research Council of Canada (NSERC) \
\end{minipage}\\
\makebox[3ex]{$^{ I}$}
\begin{minipage}[t]{14cm}
 supported by RF Presidential grant N 3042.2014.2 for the Leading Scientific Schools and by
 the Russian Ministry of Education and Science through its grant for Scientific Research on
 High Energy Physics\
\end{minipage}\\
\makebox[3ex]{$^{ J}$}
\begin{minipage}[t]{14cm}
 supported by the Israel Science Foundation\
\end{minipage}\\
\vspace{30em} \pagebreak[4]


\makebox[3ex]{$^{ a}$}
\begin{minipage}[t]{14cm}
also funded by Max Planck Institute for Physics, Munich, Germany\
\end{minipage}\\
\makebox[3ex]{$^{ b}$}
\begin{minipage}[t]{14cm}
also funded by Max Planck Institute for Physics, Munich, Germany, now at Sri Guru Granth Sahib World University, Fatehgarh Sahib\
\end{minipage}\\
\makebox[3ex]{$^{ c}$}
\begin{minipage}[t]{14cm}
also at Agensi Nuklear Malaysia, 43000 Kajang, Bangi, Malaysia\
\end{minipage}\\
\makebox[3ex]{$^{ d}$}
\begin{minipage}[t]{14cm}
also at Cracow University of Technology, Faculty of Physics, Mathematics and Applied Computer Science, Poland\
\end{minipage}\\
\makebox[3ex]{$^{ e}$}
\begin{minipage}[t]{14cm}
partially supported by the Polish National Science Centre projects DEC-2011/01/B/ST2/03643 and DEC-2011/03/B/ST2/00220\
\end{minipage}\\
\makebox[3ex]{$^{ f}$}
\begin{minipage}[t]{14cm}
now at Rockefeller University, New York, NY 10065, USA\
\end{minipage}\\
\makebox[3ex]{$^{ g}$}
\begin{minipage}[t]{14cm}
now at University of Liverpool, United Kingdom\
\end{minipage}\\
\makebox[3ex]{$^{ h}$}
\begin{minipage}[t]{14cm}
also affiliated with University College London, UK\
\end{minipage}\\
\makebox[3ex]{$^{ i}$}
\begin{minipage}[t]{14cm}
supported by the Alexander von Humboldt Foundation\
\end{minipage}\\
\makebox[3ex]{$^{ j}$}
\begin{minipage}[t]{14cm}
now at CERN, Geneva, Switzerland\
\end{minipage}\\
\makebox[3ex]{$^{ k}$}
\begin{minipage}[t]{14cm}
also at Institute of Theoretical and Experimental Physics, Moscow, Russia\
\end{minipage}\\
\makebox[3ex]{$^{ l}$}
\begin{minipage}[t]{14cm}
Alexander von Humboldt Professor; also at DESY and University of Oxford\
\end{minipage}\\
\makebox[3ex]{$^{ m}$}
\begin{minipage}[t]{14cm}
also at University of Tokyo, Japan\
\end{minipage}\\
\makebox[3ex]{$^{ n}$}
\begin{minipage}[t]{14cm}
now at Kobe University, Japan\
\end{minipage}\\
\makebox[3ex]{$^{ o}$}
\begin{minipage}[t]{14cm}
supported by DESY, Germany\
\end{minipage}\\
\makebox[3ex]{$^{ p}$}
\begin{minipage}[t]{14cm}
member of National Technical University of Ukraine, Kyiv Polytechnic Institute, Kyiv, Ukraine\
\end{minipage}\\
\makebox[3ex]{$^{ q}$}
\begin{minipage}[t]{14cm}
now at DESY ATLAS group\
\end{minipage}\\
\makebox[3ex]{$^{ r}$}
\begin{minipage}[t]{14cm}
member of National University of Kyiv - Mohyla Academy, Kyiv, Ukraine\
\end{minipage}\\
\makebox[3ex]{$^{ s}$}
\begin{minipage}[t]{14cm}
now at BNL, USA\
\end{minipage}\\
\makebox[3ex]{$^{ t}$}
\begin{minipage}[t]{14cm}
now at LNF, Frascati, Italy\
\end{minipage}\\
\makebox[3ex]{$^{ u}$}
\begin{minipage}[t]{14cm}
also at Max Planck Institute for Physics, Munich, Germany, External Scientific Member\
\end{minipage}\\
\makebox[3ex]{$^{ v}$}
\begin{minipage}[t]{14cm}
now at Tokyo Metropolitan University, Japan\
\end{minipage}\\
\makebox[3ex]{$^{ w}$}
\begin{minipage}[t]{14cm}
also supported by DESY\
\end{minipage}\\
\makebox[3ex]{$^{ x}$}
\begin{minipage}[t]{14cm}
also at \L\'{o}d\'{z} University, Poland\
\end{minipage}\\
\makebox[3ex]{$^{ y}$}
\begin{minipage}[t]{14cm}
member of \L\'{o}d\'{z} University, Poland\
\end{minipage}\\
\makebox[3ex]{$^{ z}$}
\begin{minipage}[t]{14cm}
now at Polish Air Force Academy in Deblin\
\end{minipage}\\
\makebox[3ex]{$^{aa}$}
\begin{minipage}[t]{14cm}
deceased\
\end{minipage}\\
\pagebreak[4]
}

%
\pagenumbering{arabic}
\pagestyle{plain}
\section{Introduction}
Deep inelastic scattering (DIS) in $e^{\pm}p$ collisions is an important tool to investigate the structure of the proton and to test different theoretical approaches to solving quantum chromodynamics (QCD).
At low virtuality, $Q^2$, the inclusive $e^\pm p$  DIS cross section
can be expressed in
terms of the two structure functions, $F_2$ and $F_L$, as
\begin{equation}
\frac{d^2\sigma^{e^\pm p}}{dxdQ^2} = \frac{2\pi\alpha^2Y_+}{xQ^4}
\left[F_2(x,Q^2) - \frac{y^2}{Y_+} F_L(x,Q^2)\right]
= \frac{2\pi\alpha^2Y_+}{xQ^4} \; \tilde{\sigma} (x,Q^2,y), 
\label{eq:disnc-xsec}
\end{equation}
where $\alpha$ is the fine structure constant, $x$ is the Bjorken
scaling variable, $y$ is the inelasticity and $Y_+=1+(1-y)^2$. The
quantity $\tilde{\sigma}$ is referred to as the reduced cross
section. 
The kinematic variables are related via $Q^2=xys$, where $\sqrt{s}$
is the $ep$ centre-of-mass energy.

The magnitude of $F_L$ is proportional to the absorption cross section
for longitudinally polarised virtual photons by protons,
$F_L\propto \sigma_L$, while $F_2$ includes also the absorption cross section for
transversely polarised virtual photons,
$F_2\propto(\sigma_T+\sigma_L)$. At low values of $x$, the ratio
$R=F_L/(F_2-F_L)\approx\sigma_L/\sigma_T$ gives the relative
strengths of the two components.

HERA measurements~\cite{herapdf1} of $\tilde{\sigma}$ provide the strongest
constraints on the proton parton distribution functions (PDFs) at low
$x$. Within the DGLAP formalism~\cite{Art:dokshitzer,*Art:gribov,*Art:gribov2,*Art:altarelli}, $F_2$ at low $x$ is dominated by the
quark sea distributions while the scaling violations of $F_2$
reflect the gluon distribution.

The value of $F_L$ is zero in zeroth-order perturbative QCD.
Contributions to $F_L$ arise from higher-order processes
initiated by quarks and gluons in the proton.
At low $Q^2$ and $x$,
the gluon-initiated processes dominate and $F_L$ is directly correlated
to  the gluon density.
The description of the $F_L$ contribution 
is the theoretically  most challenging part in the QCD analysis (PDF extraction) 
of the HERA reduced cross-section data.
The various available model calculations \cite{Lai2010,Martin2009,Gluck2008, Jimenez2009,Ball2013,Forte2010,Alekhin} differ sizeably in their predictions of $F_L$, in particular at low $Q^2$.

This paper presents new measurements of the reduced $e^{+}p$ cross sections up to large values of $y$ using the ZEUS detector.  The data were taken at different 
centre-of-mass energies, $\sqrt{s}=318, 251$ and  $225$~GeV, allowing for measurements at fixed $(x,Q^2)$ but different values of $y$.  This allows $F_2$ to be decoupled  from $F_L$, 
thus providing direct sensitivity to the gluon density.  The variation of $\sqrt{s}$ was achieved by varying the proton beam energy, $E_p$, while keeping the electron\footnote{Here and in the following the term electron denotes generically both the electron and the positron.} beam energy constant, $E_e=27.5$\gev. 
The data were collected in 2006 and 2007 with $E_p=920$, $575$ and $460$\gev, referred to respectively as the
high- (HER), medium- (MER) and low-energy-running (LER) samples.
The corresponding integrated luminosities of the HER, MER and LER 
samples are $44.5$, $7.1$ and $13.9~{\rm pb}^{-1}$, respectively.   
The cross-section measurements at fixed $(x,Q^2)$ and different $y$ were used to extract $F_2$ and $F_L$ separately, as has been previously done in
fixed-target experiments~\cite{Aubert:1982ts,Benvenuti:1989rh,Whitlow:1990gk,Arneodo:1996qe}, 
and recently by the ZEUS \cite{zeus-fl} and H1 \cite{Aaron:2008tx,fl-h1-2,fl-h1-3,fl-h1-diff} collaborations at HERA.
For the results reported in this paper, high-precision data taken with an $820$~GeV proton beam \cite{epj:c21:443} 
were also included in the extraction. This data set, referred to as the ZEUS97 sample, has an integrated luminosity of $30.0~{\rm pb}^{-1}$.

The data described here supersede those in the previous publication~\cite{zeus-fl}.
The main improvements with respect to the previous analysis are an extension of the kinematic coverage of the measurements to lower values of $Q^2$,
improved analysis techniques and a better understanding of systematic uncertainties, leading to more accurate and more precise measurements 
of the reduced cross sections. 
%
\section{The ZEUS detector}
A detailed description of the ZEUS detector can be found elsewhere~\cite{zeus:1993:bluebook}. A brief outline 
of the components most relevant for this analysis is given below.

In the kinematic range of the analysis, charged particles were tracked in the central tracking detector
(CTD)~\citeCTD and the microvertex detector (MVD)~\citeMVD. These components operated in a magnetic 
field of $1.43\Tesla$ provided by a thin superconducting solenoid. The CTD drift chamber covered the 
polar-angle region\footnote{The ZEUS coordinate system is a right-handed Cartesian system, with the $Z$ axis pointing in the proton beam direction, referred to as the ``forward direction'', and the $X$ axis pointing towards the centre of HERA.  The coordinate origin is at the centre of the CTD.}
\mbox{$15^\circ<\theta<164^\circ$}. The MVD silicon tracker consisted of a barrel (BMVD) and a forward
(FMVD) section. The BMVD provided polar-angle coverage for tracks from
$30^\circ$ to $150^\circ$. 
The FMVD extended the polar-angle coverage in the forward region to $7^\circ$. For CTD-MVD tracks that pass through all nine CTD superlayers, the momentum resolution was $\sigma(p_T )/p_T = 0.0029p_T\oplus0.0081\oplus0.0012/p_T$, with $p_T$ in GeV.

The high-resolution uranium--scintillator calorimeter (CAL)~\citeCAL consisted of three parts: 
the forward (FCAL), the barrel (BCAL) and the rear (RCAL) calorimeters.
Each part was subdivided transversely into towers and longitudinally into one electromagnetic
section (EMC) and either one (in RCAL) or two (in BCAL and FCAL) hadronic sections (HAC).
The smallest subdivision of the calorimeter was called a cell. The CAL energy resolutions, as measured 
under test-beam conditions, were $\sigma(E)/E=0.18/\sqrt{E}$ for electrons and $\sigma(E)/E=0.35/\sqrt{E}$ 
for hadrons, with $E$ in GeV.

The rear hadron-electron separator (RHES)~\cite{Dwurazny:1988fj} consisted of a layer of approximately $10\,000$ 
$(3\times 3\,{\rm cm^2})$ 
silicon-pad detectors inserted in the RCAL at a depth 
of approximately 3 radiation 
lengths. The small-angle rear tracking detector (SRTD)~\cite{Bamberger:1997fg} was attached to the
 front face of the RCAL and consisted of two planes 
of scintillator strips.
The detector covers the total area of $68 \times68$~cm$^2$, with a
$20\times20$~cm$^2$ cutout in the centre for the beam-pipe.  The
polar-angle coverage is $162^\circ<\theta<176^\circ$, with full
acceptance for $167^\circ<\theta<174.5^\circ$.

A small tungsten--scintillator calorimeter located approximately 6 m from the interaction
point in the rear direction is referred to as the 6m Tagger \cite{thesis:gosau,*thesis:schroder}. For scattered
electrons at small scattering angles in the energy range from 4.1 to 7.4 GeV, the acceptance was very close to unity
with very high purity.

The luminosity was measured using the Bethe--Heitler reaction $ep\,\rightarrow\, e\gamma p$ by a luminosity detector which
consisted of independent lead--scintillator calorimeter\citePCAL and magnetic spectrometer\cite{Helbich:2005qf} systems. 
The fractional systematic uncertainty on the measured luminosity was $1.8\%$\cite{nimaAdamczyk},
composed of a correlated uncertainty of $1.5\%$ for the HER, LER and MER data sets 
and an additional $1\%$ uncorrelated uncertainty for each data set.  
The ZEUS97 data set has an independent luminosity uncertainty of $1.5\%$.  
%
%
\section{Monte Carlo samples}
The DIS signal processes were
simulated using the {\sc Djangoh 1.6} Monte Carlo (MC) model~\cite{django} 
with the CTEQ5D~\cite{Art:cteq5l} parameterisation of the proton PDF.
The hadronic final state of the {\sc Djangoh} MC was 
simulated using the colour-dipole model of {\sc Ariadne 4.12}~\cite{Sjostrand:1985ys}.
Photoproduction events ($Q^2<1.5$ GeV$^2$), which are the largest background for the measurement, were simulated using
the {\sc Pythia 6.221}~\cite{Art:pythia, *Art:pythia2} MC model in
conjunction with the GRV-G-96 HO~\cite{Art:grvgloTN} parameterisation
of the photon PDF and with the CTEQ5D parameterisation of the proton PDF. 
Additional background components that were considered
were elastic QED Compton (QEDC) scattering and mis-reconstructed low-$Q^2$ DIS events, simulated using
the {\sc Grape-Compton}~\cite{Abe:2000cv} and {\sc Djangoh 1.6} MC models,
respectively. 
For all {\sc Djangoh}- and {\sc Pythia}-generated samples the 
Lund string model of {\sc Jetset 7.4}~\cite{Sjostrand:1993yb} was used for the hadronisation. 
The ZEUS detector response was simulated using a program based on 
{\sc Geant 3.21}~\cite{unp:geant}.
The generated events were passed through the detector simulation, subjected to the same
trigger requirements as the data and processed by the same reconstruction programs.

The {\sc Djangoh} and {\sc Pythia} 
samples included a diffractive component
and first-order electroweak corrections. The diffractive and non-diffractive
components of the {\sc Djangoh} sample were scaled (reweighted) to improve the description
of the $\eta_{\rm max}$ distribution, where $\eta_{\rm max}$ is equal to the
pseudorapidity of the most forward CAL energy deposit.  

The electroweak corrections were simulated using
the {\sc Heracles 4.6}~\cite{django,heracles4.6} MC model.
Their uncertainty 
was evaluated by comparing the predictions from {\sc Heracles} to the higher-order predictions
from {\sc Hector 1.0}~\cite{Arbuzov:1995id}.  The predictions were found to
agree within $0.5\%$ and the remaining uncertainty was not included in the systematic uncertainties on the cross sections. 

The predicted photoproduction backgrounds
were compared with photoproduction data selected using the 6m-Tagger 
and a photoproduction-enriched sample, separately for all three data sets, HER, LER and MER\cite{thesis:prabhdeep}.  
An event weight was determined for this background as a function of the candidate electron energy to bring the MC in agreement with the data. 
The same weighting function was found to suffice for all data sets. 
The weight reduced from $1.2$ at electron-candidate energy of 6~GeV to $1.0$ at electron-candidate energy of 10~GeV. 
A remaining uncertainty of $\pm10\%$ in the photoproduction background was accounted for in the systematic uncertainties.
\section{Vertex distribution}
Events were separately analysed for two different vertex regions: 
the central-vertex region, $|Z_{\rm vtx}| < 30$~cm, and the shifted-vertex region, $30 < Z_{\rm vtx} < 100$~cm. The 
latter region contained fewer events, but the scattered electrons from these events had a greatly increased geometrical acceptance at low $Q^2$.

The distribution of event vertices depended on the bunch structure of the proton and electron beams. 
Primary bunches were separated by $96$~ns, whereas secondary electron bunches occurred at $2$~ns intervals 
and secondary proton bunches occurred at $4.8$~ns intervals. 
Assuming that one of the interacting particles was in a primary bunch, nine separate peaks of
the vertex distribution along the $Z$ direction, within $\pm100$~cm of the nominal interaction point, were produced.  

\begin{figure}[htbp]
\begin{center}
\vspace{-2cm}
\includegraphics[scale=0.85]{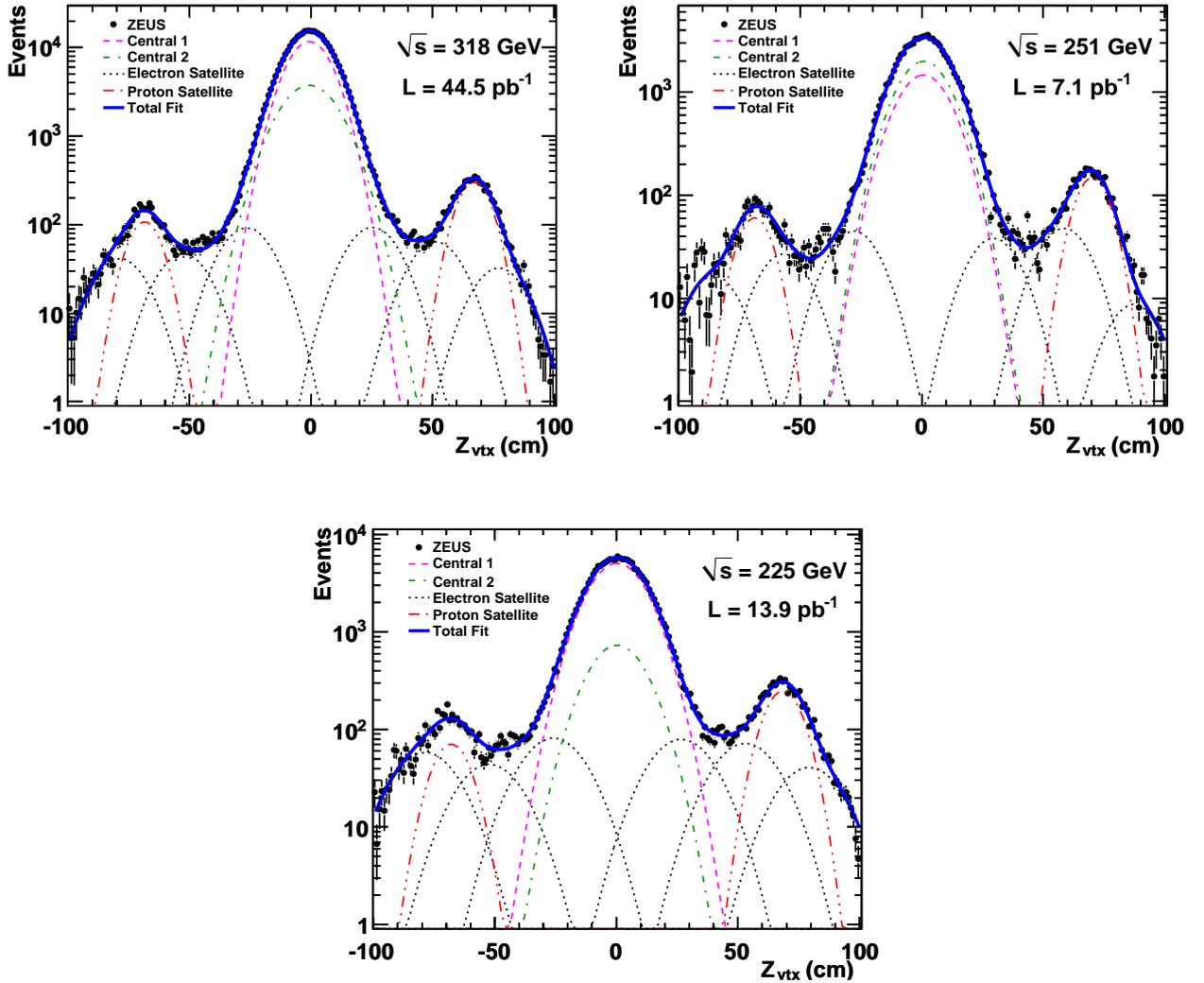}
\caption{\it Efficiency-corrected $Z_{\rm vtx}$ distribution for the HER, MER and LER data sets.
A sum of several contributions is fitted to the data: 6 Gaussian functions representing  electron satellite peaks, 2 Gaussians representing proton satellite peaks and 2 Gaussians for central vertex events. 
The fitted individual contributions are shown (dotted, dash-dotted and dashed curves) 
as well as the total fit (solid curve).
\label{fig:zvtx}}
\end{center}
\end{figure}

A dedicated analysis of the vertex distributions using separate data sets was performed, leading to a parameterisation with 
10 Gaussian functions, two Gaussian functions being required 
to fit the central peak~\cite{thesis:prabhdeep}.  
The obtained efficiency-corrected $Z_{\rm vtx}$ distributions for the HER, MER and LER samples 
are shown in Fig.~\ref{fig:zvtx}. 

The vertex distributions of the MC samples
were accurately tuned to match the distributions in the data. This was particularly crucial
to obtain proper acceptance corrections for the events selected in the two different vertex
regions.  The underlying vertex distributions were extracted using a number of different data selections, 
leading to the determination of an increased uncorrelated normalisation uncertainty for the shifted-vertex region of  $3.0$\%.
%
%
%
\section{Event reconstruction and selection}

The event kinematics were evaluated based on the reconstruction of the scattered electron~\cite{emethod} using
\begin{equation}
y_e = 1 - \frac{E^{\prime}_e}{2E_e}\left(1-\cos\theta_e\right),
\label{eq:ye}
\end{equation}
\begin{equation}
Q_e^2 = 2E^{\prime}_eE_e\left(1+\cos\theta_e\right),
\label{eq:Q2}
\end{equation}
where $\theta_e$ and $E^{\prime}_e$ are the polar angle and energy of the scattered electron, respectively.

Electrons were identified using a neural network based on the moments of the three-dimensional shower profile of clusters found in 
the CAL\cite{Sinkus:1996ch,*Abramowicz:1995zi}. The quantity $E^{\prime}_e$ was reconstructed using the CAL and $\theta_e$ was determined 
using the reconstructed interaction vertex and scattered-electron position in the SRTD or, if outside the SRTD acceptance, 
in the RHES. In less than $2\%$ of events, $\theta_e$ could not be determined in this way; such events were rejected.

The quantity $\delta \equiv \sum_{i}(E-p_Z)_i$ was used both in the trigger and in the offline analysis.
The sum runs over all CAL energy deposits. Conservation of energy, $E$, and longitudinal momentum, $p_Z$, implies that $\delta \approx 2E_e = 55$\gev.  
Undetected particles that escape through the forward beam-pipe hole 
contribute negligibly to~$\delta$. Undetected particles that escape through the rear beam-pipe hole, such as the final-state electron 
in a photoproduction event, cause a substantial reduction in $\delta$. Events not originating from $ep$ collisions often exhibit a very large $\delta$. 

A three-level trigger system was used to select events online~\cite{zeus:1993:bluebook,epj:c1:109,nim:a355:278,Gttsmith1}. 
A dedicated trigger was developed providing high efficiency for high-$y$ events~\cite{thesis:shima}.  
The trigger required an event to have $\delta$ $> 30$\gev\ and either an electron candidate with $E^{\prime}$ $>$ 4 GeV in the RCAL outside a 30 $\times$ 30 cm$^2$ box centred around the beam-pipe, or a $\delta^{\theta<165^{\circ}}$ $>$ 20 GeV, where $\delta^{\theta<165^{\circ}}$ denotes $\delta$ calculated only from the CAL energy deposits at polar angles less than 165$^{\circ}$.  The electron energy requirement was reduced to $E^{\prime}>2$~GeV for the MER and LER data runs.

Events were selected offline if:
\begin{itemize}
\item $42 < \delta < 65$\gev;
\item the reconstructed interaction vertex fulfilled $|Z_{\rm vtx}|<30$~cm for the central- and \mbox{$30<Z_{\rm vtx}<100$~cm} for the shifted-vertex region;
\item the energy of the electron candidate satisfied $E^{\prime}_e>6$\gev;\item the event topology was not compatible with an elastic QEDC scattering event, 
which has an event signature of two, and only two, electromagnetic deposits back to back in azimuthal angle;
\item the event timing was consistent with that of an $ep$ interaction;\item $y_e < 0.95$ and $y_{\rm JB} > 0.05$, where $y_{\rm JB}$ is the Jacquet--Blondel estimator~\cite{proc:epfacility:19
79:391} of $y$;\item $p_{T,h}/p_{T,e}>0.3$, where $p_{T,h}$ and $p_{T,e}$ refer to the transverse momentum of the hadronic system and electron candidate, respectively. 
\end{itemize}

The projected path of the electron candidate was required to:
\begin{itemize}\item exit the CTD at a radius $>20$~cm and hence traverse the MVD fiducial volume (typically three active layers) and at least four active layers of the CTD;
\item enter the RCAL at a radius $<135$~cm, to ensure full energy containment in the RCAL.
\end{itemize}

Hit information from the MVD and CTD was used to validate electron candidates~\cite{thesis:shima}. The procedure was based 
on the ratios of the number of observed to the maximum number of possible hits in the MVD and CTD along the track trajectory, 
denoted $f_{\rm hit}^{\rm MVD}$ and $f_{\rm hit}^{\rm CTD}$.  The requirements on the ratios were $f_{\rm hit}^{\rm MVD}>0.45$ and $f_{\rm hit}^{\rm CTD}>0.6$.
This method was used to increase the polar-angle acceptance compared to the regular tracking capability of the MVD+CTD~\cite{thesis:prabhdeep}.
The efficiency of this selection was studied in data and MC using specially selected data sets.  A reweighting of the MC for the central-vertex event sample was found to be necessary for electrons at small scattering angles, with a maximum weight of $1.06$ for the smallest angles used in this analysis.  The uncertainty on this reweighting procedure was included in the systematic uncertainties.  No reweighting was necessary for the shifted-vertex events, as the efficiency of the selection is high and uniform for this sample.

\begin{figure}[htbp]
\begin{center}
\vspace{-2cm}
\includegraphics[scale=0.85]{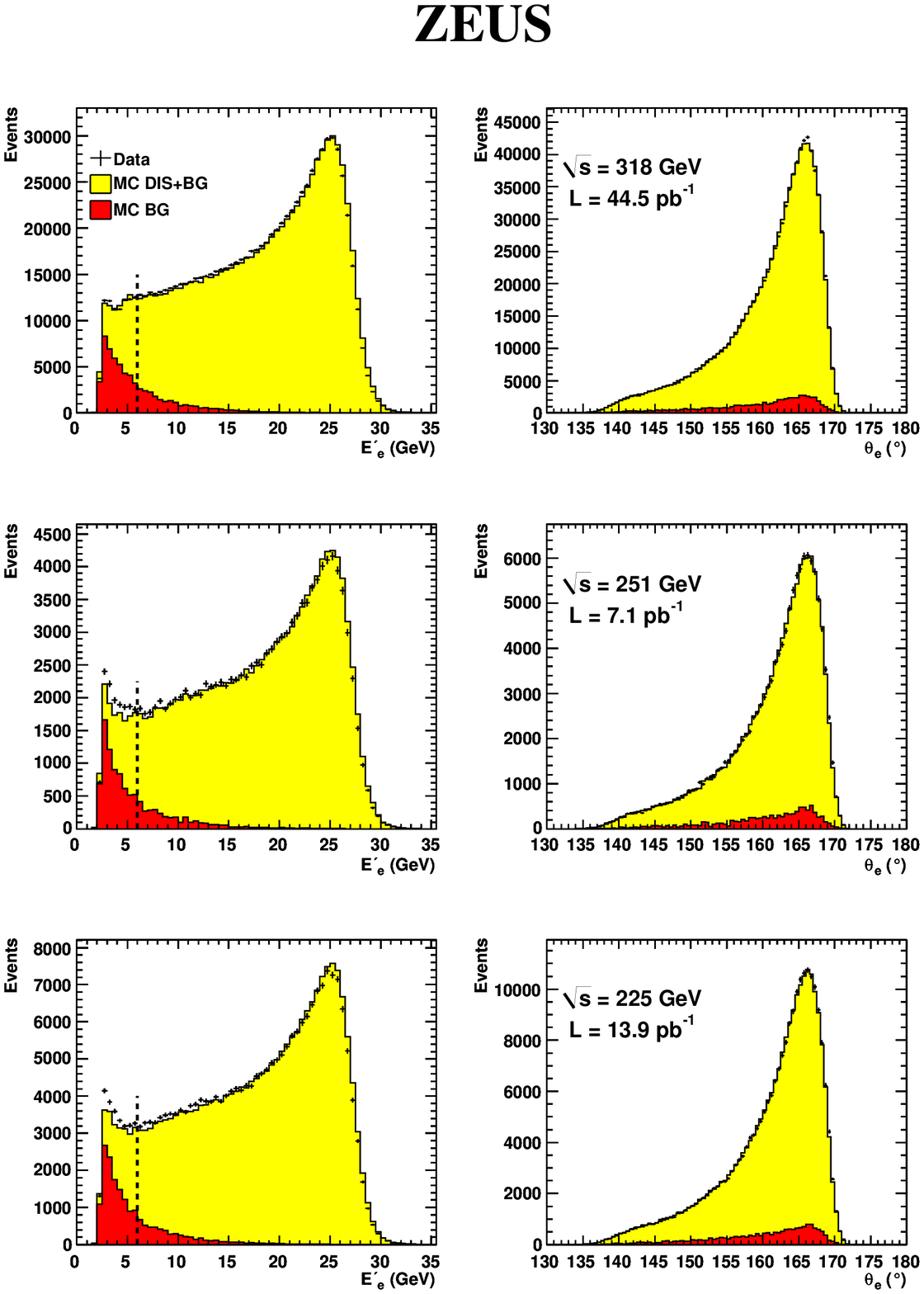}
\caption{\it Detector-level distributions of the variables $E^{\prime}_e$ and $\theta_e$ for the $|Z_{\rm vtx}|<30$ cm region for the HER, MER and LER data sets, compared to the combined MC predictions (MC DIS+BG).
The background-only MC is labelled MC BG. The vertical dashed line indicates the lower limit on the scattered-electron energy used in this analysis. The $\theta_e$ distributions are shown for $E^{\prime}_e>$ 6 \gev.
\label{fig:control_nominal}}
\end{center}
\end{figure}

\begin{figure}[htbp]
\begin{center}
\vspace{-2cm}
\includegraphics[scale=0.85]{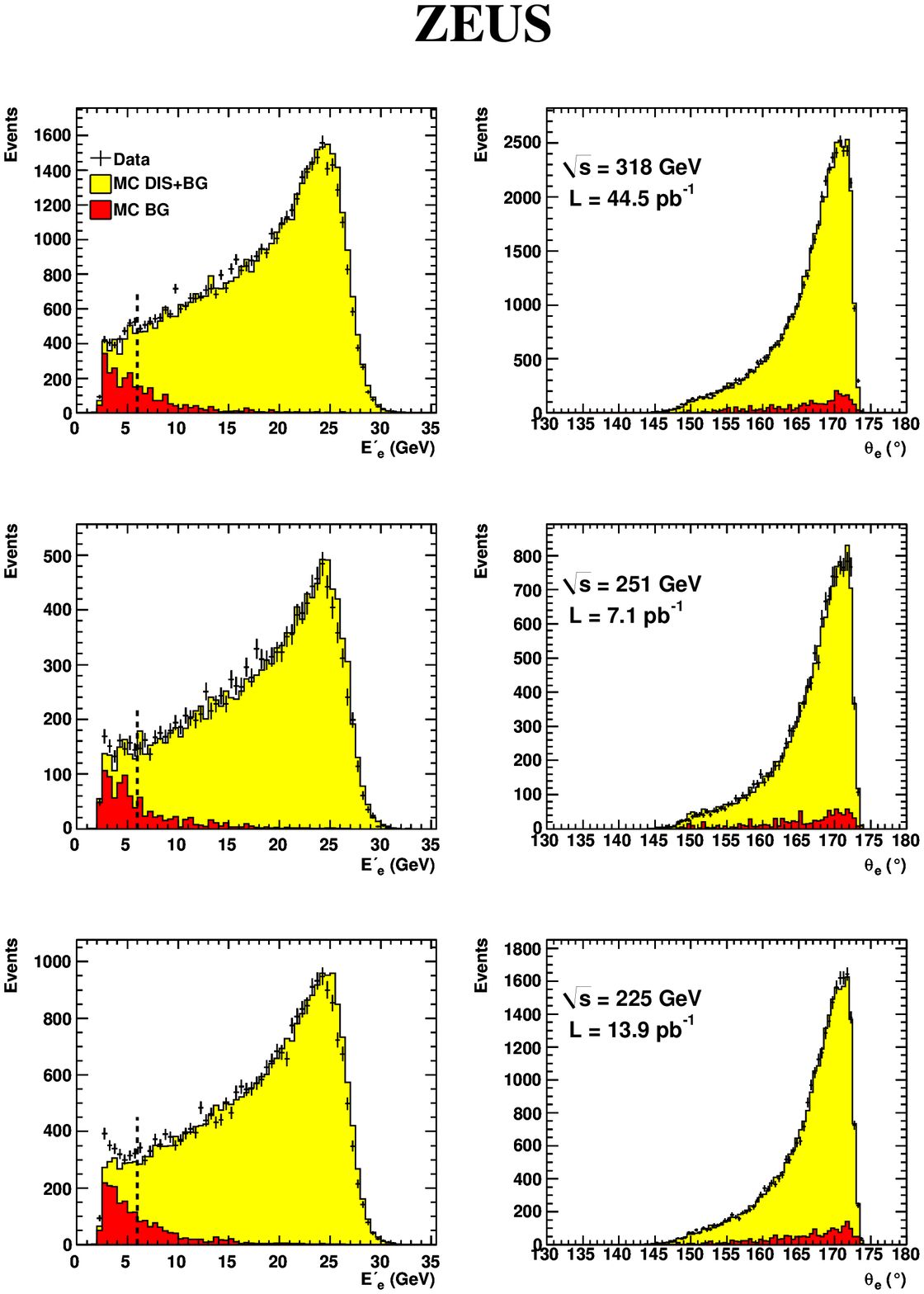}
\caption{\it Detector-level distributions of the variables $E^{\prime}_e$ and $\theta_e$ for the $30<Z_{\rm vtx}<100$ cm region for the HER, MER and LER data sets, compared to the combined MC predictions (MC DIS+BG).
The background-only MC is labelled MC BG. The vertical dashed line indicates the lower limit on the scattered-electron energy used in this analysis. The $\theta_e$ distributions are shown for $E^{\prime}_e>$ 6 \gev.
\label{fig:control_shifted}}
\end{center}
\end{figure}

Figures~\ref{fig:control_nominal} and~\ref{fig:control_shifted} show the distributions of the variables $E^{\prime}_e$
and $\theta_e$ for the HER, MER and LER data sets compared to the combined
detector-level predictions from the MC models, for the central- and shifted-vertex events.
In the region $E_e^{\prime} > 6$ GeV, which is relevant for the analysis, the
agreement is adequate for the extraction of the cross sections. According to the MC models, the final data samples 
contained 97\% DIS signal and 3\%  background events for both the central- and shifted-vertex data samples in all data sets.
The vast majority of the background events were found at low $Q^2$ and
high $y$ and resulted from photoproduction processes. In the kinematic bin most affected the background fraction was 30\%.

\section{Cross-section measurements}

The reduced cross section at a given $(x,Q^2)$ value was evaluated from the number of events reconstructed in a bin according to
\begin{equation}
 \tilde{\sigma}(x,Q^2) =  \frac{N_\mathrm{data}-N_\mathrm{MC}^\mathrm{bg}}{N_\mathrm{MC}^\mathrm{DIS}}  \, \tilde{\sigma}_\mathrm{SM}(x,Q^2) .
 \label{eq:sigma_mes}
\end{equation}
Here $N_\mathrm{data}$, $N_\mathrm{MC}^\mathrm{bg}$ and $N_\mathrm{MC}^\mathrm{DIS}$ denote,
respectively, the number of observed events in the data and the expected
number of background and signal DIS events. 
The latter two numbers were taken from the MC simulations 
normalised to the data luminosity.  
The quantity $\tilde{\sigma}_\mathrm{SM}(x,Q^2)$ denotes the Standard Model electroweak
Born-level reduced cross section and was calculated with the  {\sc Djangoh}
simulation with electroweak corrections switched off.

\begin{figure}[htbp]
\vspace{-2cm}
\hspace{-2cm}
\includegraphics[scale=0.95]{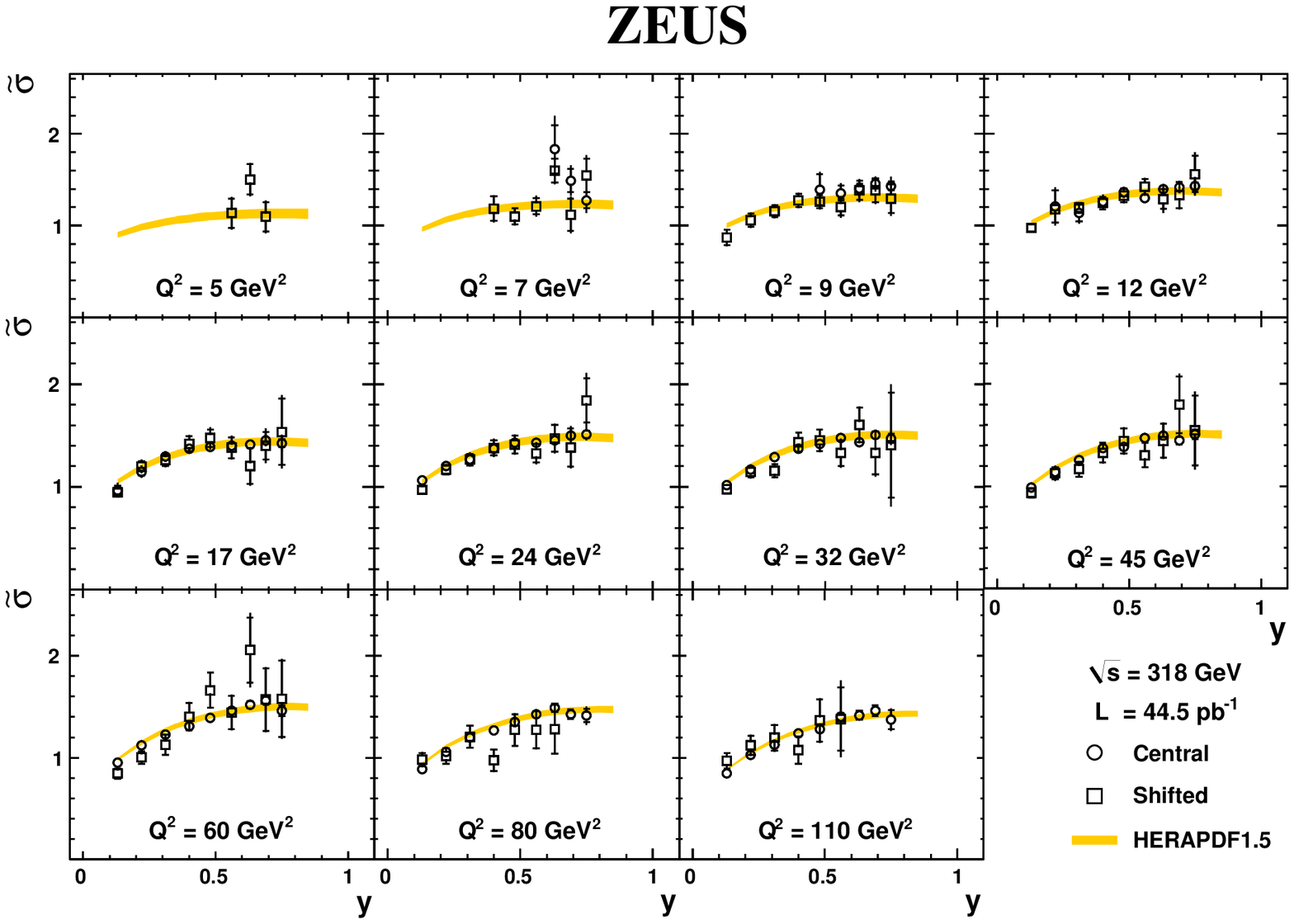}
\caption{ \it
The reduced cross sections, $\tilde{\sigma}$, for $\sqrt{s}=318$\gev, at 11 values of $Q^2$ as a function 
of $y$. The central- and shifted-vertex data results are shown as circles and squares, respectively.
The inner error bars correspond to the statistical uncertainty, while the outer error bars represent 
the statistical and systematic uncertainties 
added in quadrature. NNLO QCD predictions from HERAPDF1.5 are also shown. The bands indicate the uncertainty in the predictions.
\label{fig:HERxs}}
\end{figure}

\begin{figure}[htbp]
\vspace{-2cm}
\hspace{-1.5cm}
\includegraphics[scale=0.95]{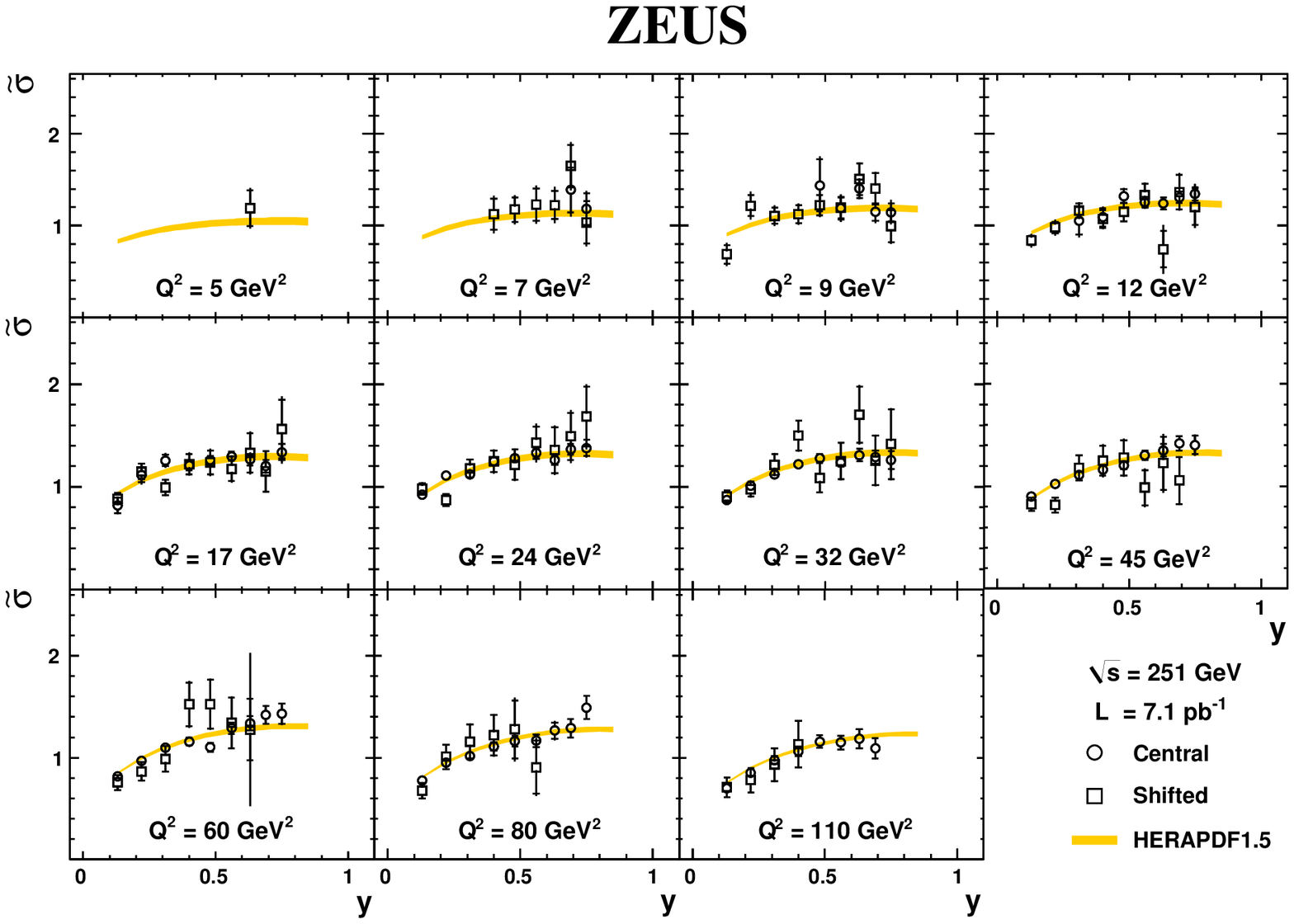}
\caption{ \it
The reduced cross sections, $\tilde{\sigma}$, for $\sqrt{s}=251$\gev, at 11 values of $Q^2$ as a function 
of $y$. The central- and shifted-vertex data results are shown as circles and squares, respectively.
The inner error bars correspond to the statistical uncertainty, while the outer error bars represent 
the statistical and systematic uncertainties 
added in quadrature. NNLO QCD predictions from HERAPDF1.5 are also shown. The bands indicate the uncertainty in the predictions.
\label{fig:MERxs}}
\end{figure}

\begin{figure}[htbp]
\vspace{-2cm}
\hspace{-2cm}
\includegraphics[scale=0.95]{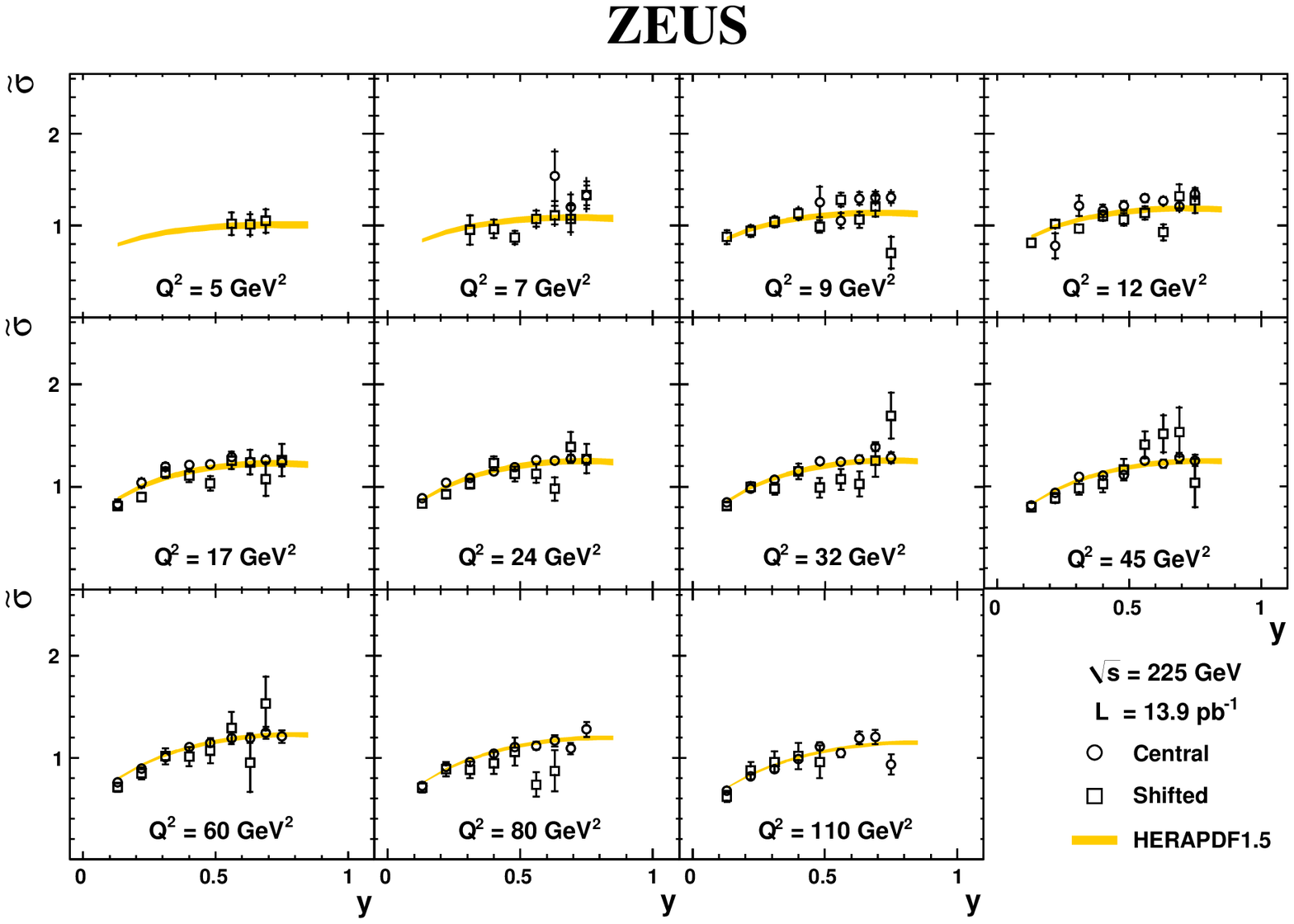}
\caption{\it
The reduced cross sections, $\tilde{\sigma}$, for $\sqrt{s}=225$\gev, at 11 values of $Q^2$ as a function 
of $y$. The central- and shifted-vertex data results are shown as circles and squares, respectively.
The inner error bars correspond to the statistical uncertainty, while the outer error bars represent 
the statistical and systematic uncertainties 
added in quadrature. NNLO QCD predictions from HERAPDF1.5 are also shown. The bands indicate the uncertainty in the predictions.
\label{fig:LERxs}}
\end{figure}

The reduced cross sections were measured according to Eq.~(\ref{eq:sigma_mes}) 
for kinematic values in the ranges $0.13\le y \le 0.75$ and $5\le Q^2\le110$\gev$^2$. 
They were measured separately for the central- and shifted-vertex regions, referred to as $\tilde{\sigma}_\mathrm{cen}$ and $\tilde{\sigma}_\mathrm{sh}$. The following $y$-bin boundaries were used for the measurement: 0.090, 0.175, 0.265, 0.355, 0.440, 0.520, 0.600, 0.660, 0.720, 0.780, together with the following $Q^2$-bin boundaries: 4.5, 6, 8, 11, 15, 20, 28, 38, 52, 70, 95, 130 \gev$^2$.
Figures~\ref{fig:HERxs}--\ref{fig:LERxs} show the obtained cross sections at fixed values of $Q^2$ as functions of $y$ for the HER, MER and LER samples, separately for
the central- and shifted-vertex regions. 
Reasonable agreement is observed between them.
The predictions based on
the HERAPDF1.5 PDF set~\cite{herapdf1.5} are also depicted
in the Figs.~\ref{fig:HERxs}--\ref{fig:LERxs} and provide a good description of the data.
The reduced cross sections are reported in Tables~\ref{tab:sigHERcen}--\ref{tab:sigLERsh} for central- and shifted-vertex regions and
are used for the extraction of structure functions.
%
%
%
\section{Systematic uncertainties}
\label{sec:syst}
A number of possible effects that could affect the cross-section measurements have been investigated.  
These are classified according to whether they affect measurements in a correlated way, can be treated as uncorrelated, or are negligibly small.  The systematic uncertainties are denoted with the symbols $\delta_{\rm source}$. The numbers in the
parentheses give an indication of the typical values observed in the cross-section bins and the maximum observed value. 
The correlated systematic uncertainties are:

\begin{itemize}
\item $\{\delta_{\gamma p}\}$, the $\pm 10\%$ normalisation uncertainty 
on the level of photoproduction background ($0.5\%, 3\%$);   

\item $\{\delta_{E_{\rm had}}\}$, the $\pm2\%$ hadronic-energy-scale uncertainty, evaluated by varying
the scale in simulated events ($0.5\%, 4\%$); 

\item $\{\delta_{\rm diff}\}$, the uncertainty on the scale factors applied to the diffractive {\sc Djangoh} component ($0.2\%, 0.5\%$);

\item $\{\delta_{\rm hits}\}$, the electron validation using CTD and MVD information ($1\%, 2\%$).

\end{itemize}

These $\delta_{\rm source}$ values
are listed in Tables~\ref{tab:sigHERcen}--\ref{tab:sigLERsh} for the 
reduced cross sections at the three different centre-of-mass energies.
The reduced cross-section changes are quoted with a sign indicating how they should be handled in a fit; positive quantities indicate an increase in the cross sections, while negative quantities indicate a decrease. The probability distributions for the cross sections are taken to be Gaussian. The quoted numbers in the table correspond to variations of one standard deviation. %

A number of systematic tests produced sizeable changes in the extracted cross sections, but should be considered as uncorrelated in fits to the cross sections.  These are:

\begin{itemize}
\item $\{\delta_{E_e}\}$, the electron-energy-scale uncertainty \cite{zeus-fl}
of $\pm 0.5\%$ for $E^{\prime}_e>20$\gev, increasing to $\pm 1.9\%$ 
at $E^{\prime}_e=6$\gev, evaluated by varying the scale
in simulated events ($1\%, 5\%$); 
\item $\{\delta_{eID}\}$, the uncertainty of the electron-finding efficiency, evaluated by 
 loosening or tightening the criterion applied to the output 
of the neural network used to select electron candidates, both for data and MC ($0.2\%, 5\%$); 
\item $\{\delta_{\rm Zvtx}\}$, the uncertainty on the $Z_{\rm vtx}$ distribution, that was evaluated by varying the event-selection criteria used in the extraction 
of the distribution ($0.5\%, 10\%$); 

\item $\{\delta_{dx},~\delta_{dy}\}$, the SRTD and HES position uncertainty 
of $\pm 2$~mm in both the horizontal and vertical directions  ($0.5\%, 4\%$).

\end{itemize}

These uncertainties are combined in quadrature for each bin and quoted under the label $\delta_{\rm unc}$. This uncertainty also includes the statistical uncertainty in the MC sample. 

Finally, a number of checks were performed with resulting cross-section variations small enough that they are considered negligible.  These include:
\begin{itemize}
\item the trigger-efficiency uncertainty;
\item the uncertainty due to the electroweak
corrections.  In a separate analysis~\cite{thesis:jason} the differential cross sections 
for radiating an initial-state photon were also extracted.  
These results are in excellent agreement with the expectations from {\sc Heracles}, adding confidence that the radiative effects are well simulated.
\end{itemize}

In addition to the uncertainties listed above, the normalisation uncertainties for the different data sets are:
\begin{itemize}
\item a correlated luminosity uncertainty of $1.5$\% for the HER, MER and LER cross sections;
\item an additional uncorrelated luminosity and vertex-distribution uncertainty of $1.0$\% for the central-vertex HER, MER and LER cross sections and $3.0$\% for the shifted-vertex cross sections.
\end{itemize}

The total systematic uncertainty, $\delta_{\rm sys}$, in each bin  
formed by adding the individual uncertainties in quadrature,
is also given in Tables~\ref{tab:sigHERcen}--\ref{tab:sigLERsh}. 
The total systematic uncertainties  in the tables do not include the normalisation uncertainty.  

\section{The ZEUS97 data set}
\label{sec:9697}
In order to increase the precision of the $F_L$ and $F_2$ extractions, the reduced NC cross sections measured by ZEUS from the $e^{+}p$ data taken 
in 1996 and 1997 with the proton beam energy $E_p$ = 820 GeV~\cite{epj:c21:443} were included in the analysis. 
The precision of these data is comparable to the HER data presented in this paper.
The binning for the ZEUS97 sample is similar but not identical to that used for the present measurement, 
so a binning correction was applied to the cross sections from the ZEUS97 sample, using the procedure introduced in an earlier publication~\cite{herapdf1}. 
An interpolation of a measurement to the required point on the $(x,Q^2)$ grid is performed 
by multiplying the measured cross section by the ratio of predicted double-differential cross sections at two different $(x,Q^2)$ points. 
For the theory calculation, the HERAPDF1.0 NLO prediction~\cite{herapdf1} was used.
Only the points which required less than $2\%$ adjustments were included in the fits to extract $F_2$ and $F_L$.

The following systematic uncertainties~\cite{epj:c21:443} on the cross sections were included in the extraction of $F_2$ and $F_L$: 
\begin{itemize}
\item $\{\tilde \delta_{\rm eID}\}$, the uncertainty in the positron-finding efficiency;
\item $\{\tilde \delta_{\rm dxdy}\}$, the uncertainty in the positron position;
\item $\{\tilde \delta_{e\theta}\}$, the uncertainty in the positron-scattering angle;
\item $\{\tilde \delta_{E_e}\}$, the uncertainty in the positron-energy scale;
\item $\{\tilde \delta_{\rm hdFC}\}$, $\{\tilde \delta_{\rm hdBC}\}$, $\{\tilde \delta_{\rm hdRC}\}$, the uncertainty in the hadronic-energy scales in FCAL, BCAL and RCAL;
\item $\{\tilde \delta_{\rm hdF}\}$, $\{\tilde \delta_{\rm Bhd}\}$, the uncertainty in the hadronic-energy flow;
\item $\{\tilde \delta_{\gamma p}\}$, the uncertainty in the photoproduction background. 
\end{itemize}

Following the prescription in the publication~\cite{epj:c21:443}, all uncertainties were treated as correlated.

The fractional systematic uncertainties after the cross-section adjustment were kept unchanged. In addition to the uncertainties mentioned above, the ZEUS97 cross sections have a $2$\% normalisation uncertainty, which resulted from  a $1.5$\% luminosity uncertainty, a $1$\% trigger uncertainty, a $1$\% uncertainty from the vertex distribution and a $0.5$\% uncertainty due to radiative corrections.

The reduced cross sections after the adjustment are given double differentially in bins of $x$ and $Q^2$ in Table~\ref{tab:sig9697}.

\section{\boldmath Extraction of $F_2$, $F_L$ and $R$}
\label{sec:extraction}

\subsection{Fit method}
\label{sec:fits}

The values of $F_L$, $F_2$ and $R=F_L/(F_2-F_L)$ were extracted by performing a fit to the reduced cross sections 
using Eq.~(\ref{eq:disnc-xsec}).
The fit was performed with the BAT package~\cite{Caldwell:2008fw}.
Prior to fitting, the HER, MER and LER cross sections were normalised to ZEUS97 data at low $y$.  This resulted in the following normalisation  factors
for the central- and shifted-vertex samples:
$c_{\rm HER,cen}=1.025\pm 0.003$; $c_{\rm MER,cen}=1.004\pm 0.007$; $c_{\rm LER,cen}=0.998\pm 0.004$; $c_{\rm HER,sh}=1.042\pm 0.012$; $c_{\rm MER,sh}=1.056\pm 0.020$; and $c_{\rm LER,sh}=1.039\pm 0.013$. 
Within the assigned global normalisation uncertainties of the respective data sets these numbers are consistent with unity.
A different cross-section binning was used for the structure-function extraction in order to cover similar $Q^2$ and $x$ ranges in each data set,  resulting in 27 $(x,Q^2)$ bins. After normalisation, the reduced cross sections from the central- and shifted-vertex regions were combined using a weighted average based on their combined statistical and uncorrelated systematic uncertainties.  This procedure, and the use of ZEUS97 data, yielded a total of 104 reduced cross sections.  
 
 The procedure to extract $F_L$, $F_2$ and $R$ is explained in the previous ZEUS publication \cite{zeus-fl}. 
 Free parameters were introduced in the fits for each of the data sets to allow the normalisation factors to vary according to Gaussian distributions with standard deviations given by the uncertainties quoted above for the HER, MER and LER data sets, respectively. The uncertainties for the central-vertex sample were used for $Q^2 \ge 17$~GeV$^2$, where these data dominated the combinations. The uncertainties for the shifted-vertex sample were used at lower $Q^2$, where they are more precise. The normalisation of the ZEUS97 data was kept fixed in the fit.  The resulting values of $F_2$ and $F_L$ therefore have an additional uncertainty resulting from the normalisation uncertainty of this data set ($2$\%).

Additional nuisance parameters were introduced in the fits to account for the correlated systematic uncertainties. 

All systematic uncertainties of the ZEUS97 data set were treated as uncorrelated with those of the HER, MER and LER samples.
A global fit was performed 
to extract $F_L$ and $F_2$ values from the reduced cross sections. 
The joint posterior probability distribution for all parameters was then extracted under different sets of conditions as described below.

\subsection{Fit results}
\label{sec:fitresults}

\begin{figure}[htbp]
\vspace{-2cm}
\hspace{-1.5cm}
\includegraphics[scale=0.93]{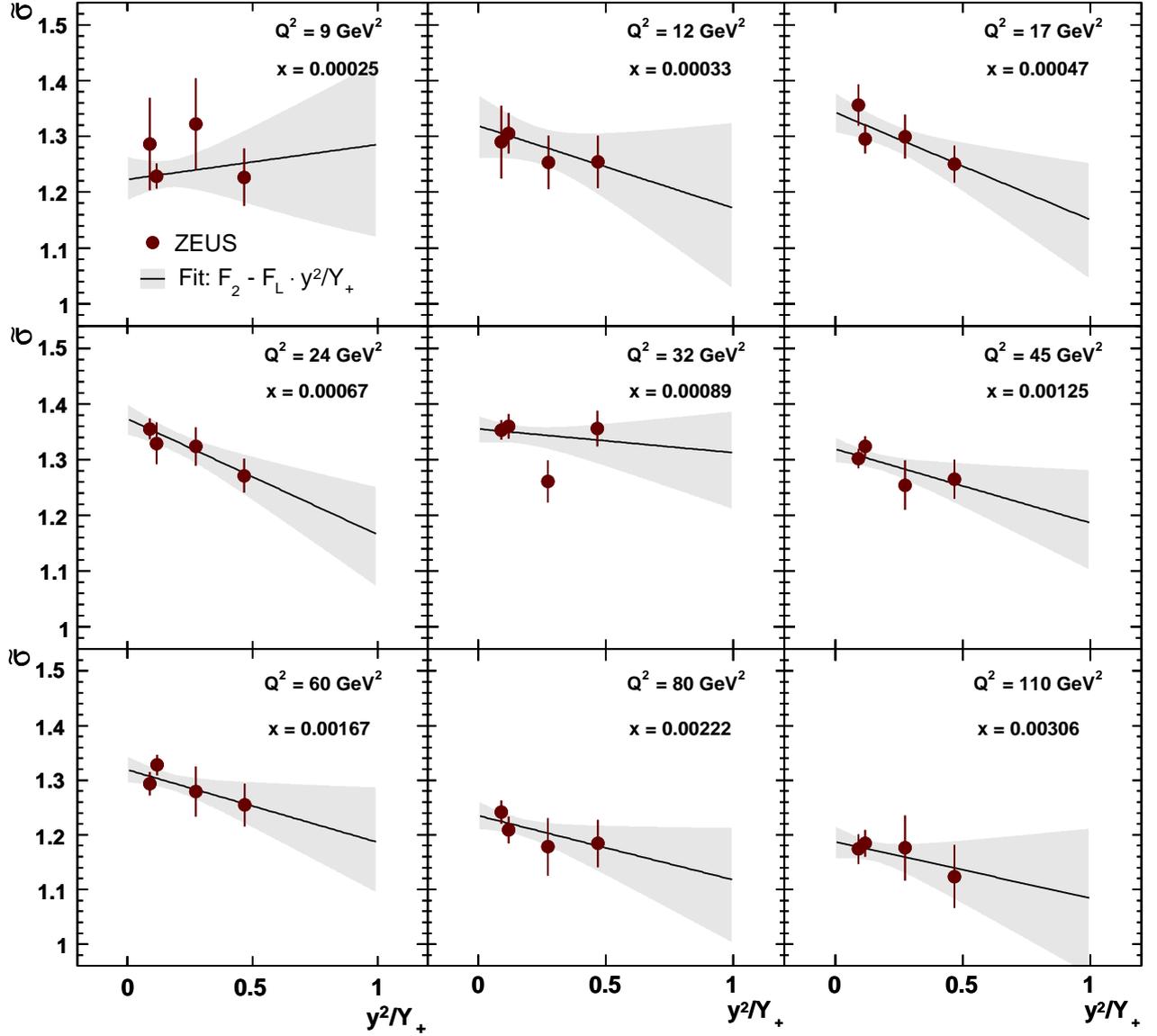}
\caption{ \it
Rosenbluth plots for the lowest-$x$ points for each of the 9 $Q^2$ values for which $F_L$ was measured. 
The reduced cross sections are shown as points, with error bars representing the combined statistical and 
systematic uncertainties. The line represents the result from the fit to the reduced cross sections (see Section~\ref{sec:extraction}), shown together with the 68$\%$ uncertainty band. 
\label{fig:rose}}
\end{figure}

Constrained and unconstrained parameter fits were performed to extract $F_2(x,Q^2)$ and $F_L(x,Q^2)$. Flat priors were used for the physics parameters within the range allowed. For the constrained fits, $0.8 < F_2 <  2$ and 0 $\leq$ $F_L$ $\leq$ $F_2$ were required. Example Rosenbluth plots~\cite{rosenbluth} 
showing the linear fit to the reduced cross sections together with a 68$\%$ uncertainty band 
are shown in Fig.~\ref{fig:rose} for the nine lowest $x$ points for each $Q^2$ bin for the unconstrained fits. 
The resulting $F_2$ and $F_L$ values for all $(x, Q^2)$ bins are given in Table \ref{tab:F2FLxQ2}.
The results given are for the parameter value at the mode of the marginalised probability distribution \cite{Caldwell:2008fw}. For each parameter, 
the narrowest $68\%$ probability interval around the marginalised mode is taken as an uncertainty. 
These ranges contain the full experimental uncertainty. Relative uncertainties as small as $2$\% were achieved for $F_2$ and (absolute) uncertainties for $F_L$ were in the range $0.1-0.2$. The results for $F_L$ found in this analysis are somewhat lower than those in the previous ZEUS analysis~\cite{zeus-fl} in the region of kinematic overlap. 
The difference compared to the previous ZEUS analysis is primarily due to improvements in the treatment of the diffractive events in the MC simulation and of the electron validation at small scattering angles.

\begin{figure}[htbp]
\vspace{-2cm}
\hspace{-1.5cm}
\includegraphics[scale=0.95]{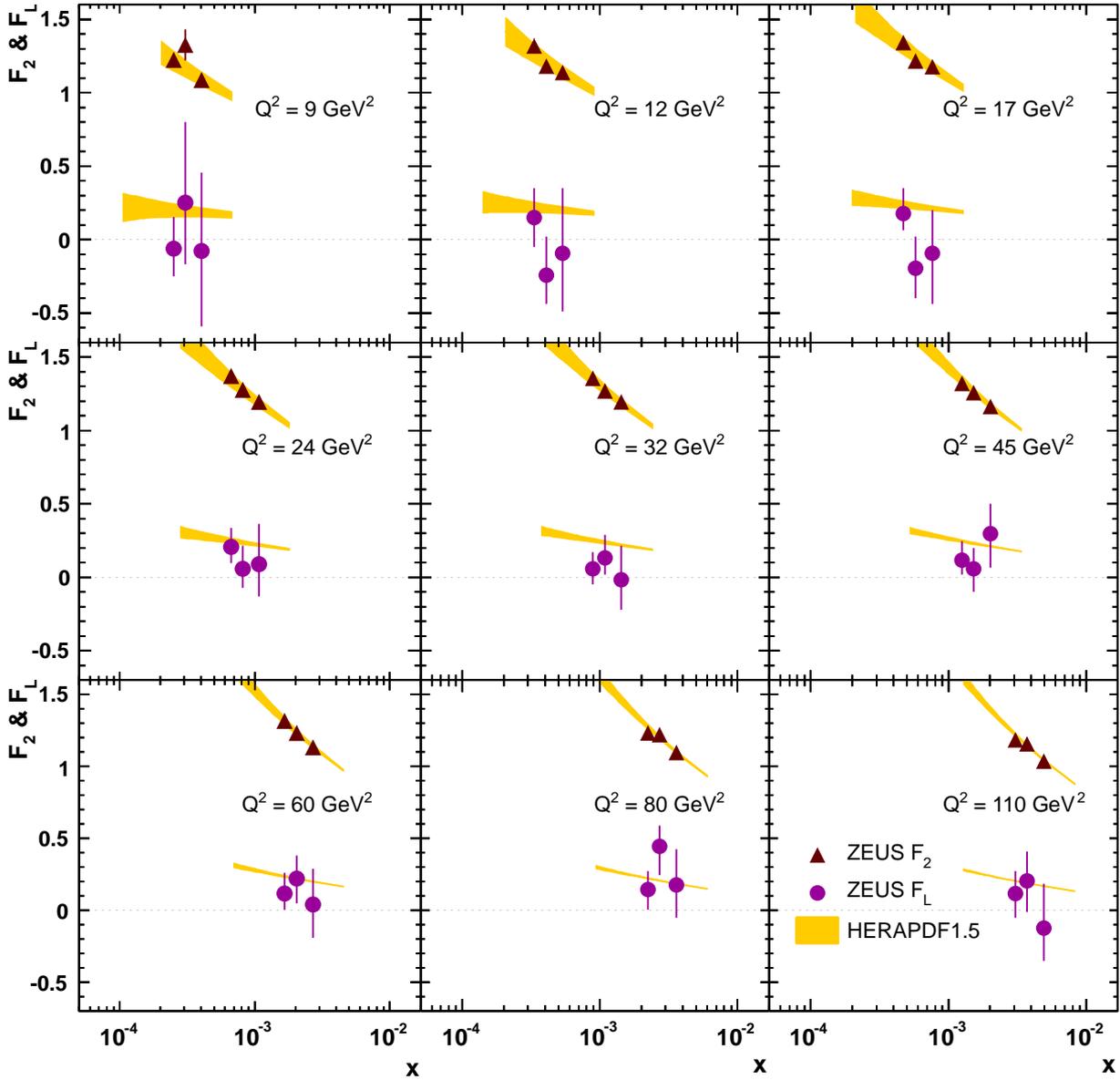}
\caption{ \it
$F_L$ and $F_2$ values as a function of $x$ for 9 values of $Q^2$, extracted from the unconstrained fit 
(see Section~\ref{sec:fitresults}). 
The error bars on the data represent the combined statistical and 
systematic uncertainties. The error bars on $F_2$ are typically smaller than the
symbols. NNLO QCD predictions from HERAPDF1.5 are also shown. The bands indicate the uncertainty in the predictions. 
\label{fig:F2FL_x}}
\end{figure}

The results from the unconstrained fit are shown in Fig.~\ref{fig:F2FL_x}  together with predictions based on the  HERAPDF1.5 expectations. 
Reasonable agreement between extracted and predicted $F_2$ and $F_L$ values is observed 
although with a tendency for the prediction to lie below the  
extracted $F_L$ values.

Further fits to the data were performed to extract $F_L(Q^2)$,  
$R(Q^2)$, and a single overall value of $R$ for the full data set. In each case, the 
same fitting procedure as described above was used, but with a reduced number 
of parameters. 

\begin{figure}[htbp]
\vspace{-2cm}
\hspace{-1.5cm}
\includegraphics[scale=0.95]{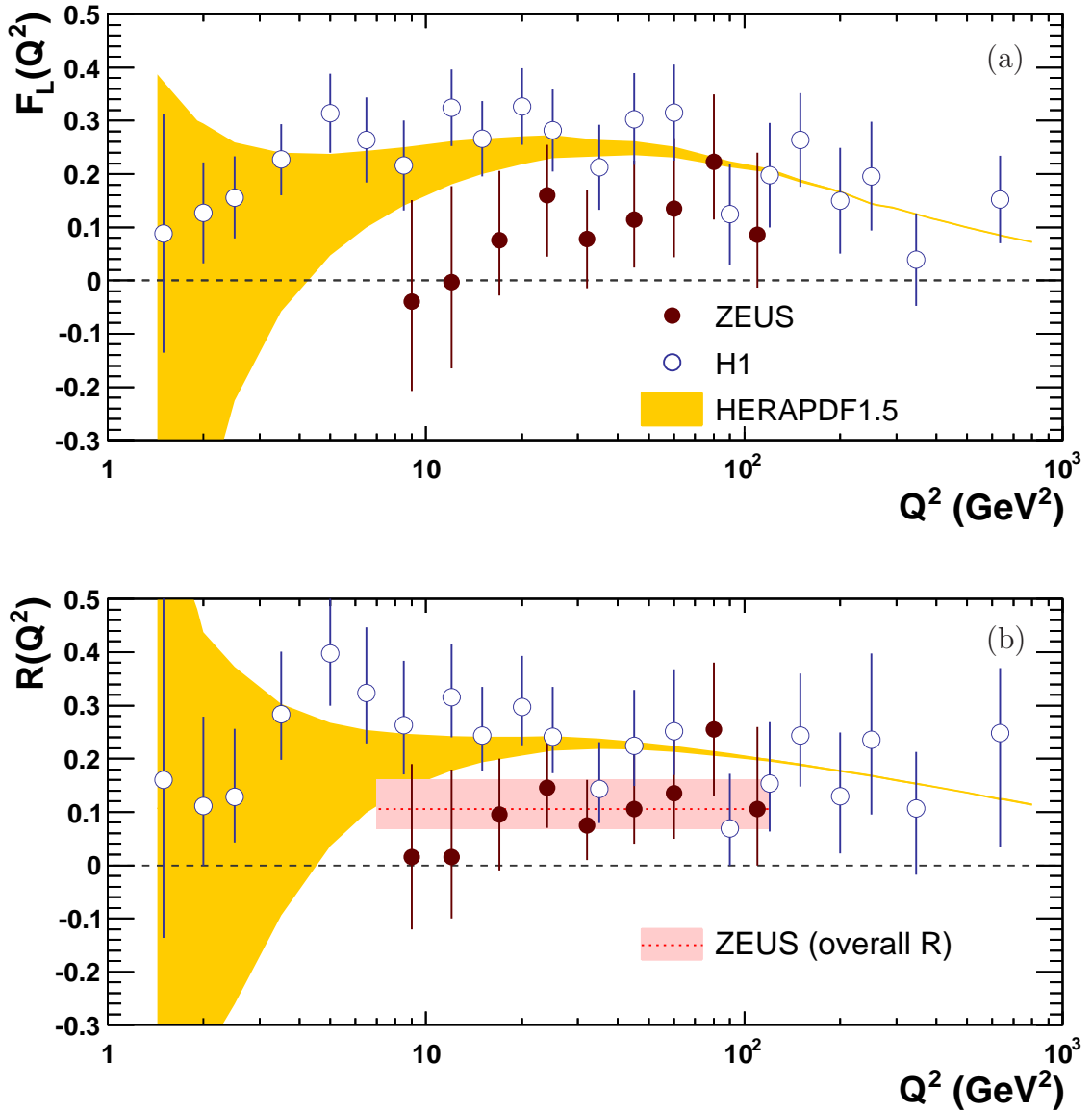}
\put(-100,430){(a)}%
\put(-100,202){(b)}%
\caption{ \it
Values of (a) $F_L$ and (b) $R$ as a function of $Q^2$. 
The ZEUS data, extracted from the unconstrained fit (see Section~\ref{sec:fitresults})
are shown as filled circles, with error bars representing the combined statistical and systematic uncertainties.
The values of $F_L$ at different $Q^2$ points are correlated, as well as the values of R.
The shaded band labelled ``ZEUS (overall R)'' represents the 68\% probability interval for the overall $R$.
The H1 data are shown as open circles with error bars representing the total uncertainties.
The ZEUS and H1 points are measured at somewhat different x values.
NNLO QCD predictions from HERAPDF1.5 are also shown. The bands indicate the uncertainty in the predictions.
\label{fig:F2FL_R_Q2}}
\end{figure}

To extract $F_L(Q^2)$, first $r(Q^2)$ was fitted, where $r = F_L/F_2$, taking a single value of $r$ for all $x$ points in the same $Q^2$ bin. The value of $F_L(Q^2)$ was then evaluated
as $F_L(x_i,Q^2) = r(Q^2)F_2(x_i,Q^2)$, where for each $Q^2$ point, $x_i$ was chosen such that $Q^2/x_i$
was constant, which for $\sqrt{s} = 225$ GeV corresponds to $y = 0.71$. Both constrained and unconstrained fits were made.  
For the constrained $R(Q^2)$ and overall $R$ fits it was required that $R(Q^2)\ge 0$ and $R\ge 0$.
The results for $F_L(Q^2)$ and $R(Q^2)$ are given in Table~\ref{tab:FLRQ2}.
The overall value of $R$ from both the unconstrained and constrained fits is $R=0.105^{+0.055}_{-0.037}$.
Figures~\ref{fig:F2FL_R_Q2}a and \ref{fig:F2FL_R_Q2}b show a comparison 
of $F_L(Q^2)$ and $R(Q^2)$ with the H1 data~\cite{fl-h1-3} and the NNLO QCD predictions based on the HERAPDF1.5. The H1 measurements generally lie above the ZEUS results. The differences were examined: taking into account the correlations between the ZEUS data points and neglecting the correlations between the H1 data points 
a $\chi^2$ of $12.2$ is obtained for $8$ degrees of freedom.  
The predictions based on HERAPDF1.5 are in reasonable agreement with both data sets.

\section{Summary}

The reduced cross sections $\tilde{\sigma}(x,Q^2)$ for $e^{+}p$ neutral current deep inelastic scattering have been measured with the ZEUS detector
at HERA, using data collected at $\sqrt{s}=318$, 
251 and 225 GeV, in the kinematic region 
$0.13\le y\le 0.75$ and $5 \le Q^2 \le 110$\gev$^2$.
The extension of the kinematic range in comparison to the previous ZEUS publication \cite{zeus-fl} was made possible with the use of shifted-vertex data.  
The new results supersede those in the previous publication.
The reduced cross sections were used 
together with those from previous ZEUS data  
collected at $\sqrt{s}=300$ GeV
to extract the proton structure functions $F_2$ and $F_L$ for 27 values of $x$ and $Q^2$. Relative uncertainties as small as $2$\% were achieved for $F_2$ and (absolute) uncertainties for $F_L$ were in the range $0.1-0.2$.  In addition, $F_L$ and the ratio, $R=F_2/(F_2-F_L)$, have also been extracted as a function of $Q^2$ together with an overall value of $R=0.105^{+0.055}_{-0.037}$.
The $F_L$ measurements reported here are lower than but compatible with those in the previous ZEUS and H1 publications and  in
reasonable agreement with theoretical predictions for $F_L$.


\section*{Acknowledgements}

We appreciate the contributions to the construction, maintenance and operation 
of the ZEUS detector made by many people who are not listed as authors.
The HERA machine group and the DESY computing staff
are especially acknowledged for their success
in providing excellent operation of the collider
and the data-analysis environment.
We thank the DESY directorate for their strong support and encouragement. We thank H.~Spiesberger for useful discussions concerning electroweak corrections. 


{
\def\bibname{\Large\bf References}
\def\refname{\Large\bf References}
\pagestyle{plain}
\ifzeusbst
  \bibliographystyle{./BiBTeX/bst/l4z_default}
\fi
\ifzdrftbst
  \bibliographystyle{./BiBTeX/bst/l4z_draft}
\fi
\ifzbstepj
  \bibliographystyle{./BiBTeX/bst/l4z_epj}
\fi
\ifzbstnp
  \bibliographystyle{./BiBTeX/bst/l4z_np}
\fi
\ifzbstpl
  \bibliographystyle{./BiBTeX/bst/l4z_pl}
\fi
{\raggedright
\bibliography{./BiBTeX/user/syn.bib,%
              ./BiBTeX/bib/l4z_articles.bib,%
              ./BiBTeX/bib/l4z_books.bib,%
              ./BiBTeX/bib/l4z_conferences.bib,%
              ./BiBTeX/bib/l4z_h1.bib,%
              ./BiBTeX/bib/l4z_misc.bib,%
              ./BiBTeX/bib/l4z_old.bib,%
              ./BiBTeX/bib/l4z_preprints.bib,%
              ./BiBTeX/bib/l4z_replaced.bib,%
              ./BiBTeX/bib/l4z_temporary.bib,%
              ./BiBTeX/bib/l4z_zeus.bib}}
}
\vfill\eject

\newpage


\begin{center}          
\scriptsize                
\begin{longtable}{|r|c|c||c|c|r|r||r|r|r|r|}             
\hline                 
$Q^2$ & $x$ & $y$ & $\tilde{\sigma}$ & $\delta_{\rm stat}$ &$\delta_{\rm sys}$ & $\delta_{\rm unc}$ & $\delta_{\gamma p}$ &$\delta_{E_{\rm had}}$ & $\delta_{\rm diff}$ & $\delta_{\rm hits}$\\  
(GeV$^2$) & & & HER & (\%) & (\%)  & (\%) & (\%) & (\%) & (\%) & (\%) \\\hline 
7 & 0.000110 & 0.63 & 1.810 & 10.05 & 17.30 & 16.70 & 0.00 & --4.09 & --0.61 & 1.83 \\
7 & 0.000100 & 0.69 & 1.470 & 5.39 & 10.06 & 9.63 & --0.97 & --1.69 & --0.56 & 2.06 \\
7 & 0.000092 & 0.75 & 1.257 & 3.45 & 10.20 & 7.36 & --3.14 & --5.86 & --0.63 & 2.29 \\
\hline                
9 & 0.000185 & 0.48 & 1.372 & 8.67 & 10.84 & 10.62 & --0.57 & --0.70 & --0.10 & 1.99 \\
9 & 0.000159 & 0.56 & 1.334 & 4.25 & 6.32 & 5.95 & --0.61 & 0.32 & --0.25 & 2.00 \\
9 & 0.000141 & 0.63 & 1.382 & 2.68 & 5.09 & 4.57 & --0.44 & --1.05 & --0.42 & 1.89 \\
\end{longtable}
\end{center}   

\newpage
\begin{center}          
\scriptsize                
\begin{longtable}{|r|c|c||c|c|r|r||r|r|r|r|}             
\hline                 
$Q^2$ & $x$ & $y$ & $\tilde{\sigma}$ & $\delta_{\rm stat}$ &$\delta_{\rm sys}$ & $\delta_{\rm unc}$ & $\delta_{\gamma p}$ &$\delta_{E_{\rm had}}$ & $\delta_{\rm diff}$ & $\delta_{\rm hits}$\\  
(GeV$^2$) & & & HER & (\%) & (\%)  & (\%) & (\%) & (\%) & (\%) & (\%) \\\hline 
5 & 0.000088 & 0.56 & 1.134 & 10.30 & 11.78 & 11.68 & --0.55 & --1.06 & --0.13 & 0.97 \\
5 & 0.000078 & 0.63 & 1.503 & 7.34 & 9.60 & 9.50 & --0.40 & 0.73 & --0.22 & 1.08 \\
5 & 0.000072 & 0.69 & 1.096 & 8.96 & 13.47 & 12.82 & --1.74 & --3.50 & --0.41 & 1.26 \\
\hline                
7 & 0.000173 & 0.40 & 1.185 & 8.02 & 9.73 & 9.67 & --0.34 & 0.44 & 0.01 & 0.87 \\
7 & 0.000144 & 0.48 & 1.101 & 6.07 & 6.50 & 6.39 & --0.19 & --0.53 & --0.04 & 1.03 \\
7 & 0.000124 & 0.56 & 1.210 & 5.40 & 7.79 & 7.68 & --0.32 & 0.72 & --0.23 & 1.06 \\
\end{longtable}
\end{center}   

\newpage
\begin{center}          
\scriptsize                
\begin{longtable}{|r|c|c||c|c|r|r||r|r|r|r|}             
\hline                 
$Q^2$ & $x$ & $y$ & $\tilde{\sigma}$ & $\delta_{\rm stat}$ &$\delta_{\rm sys}$ & $\delta_{\rm unc}$ & $\delta_{\gamma p}$ &$\delta_{E_{\rm had}}$ & $\delta_{\rm diff}$ & $\delta_{\rm hits}$\\  
(GeV$^2$) & & & MER & (\%) & (\%)  & (\%) & (\%) & (\%) & (\%) & (\%) \\\hline 
7 & 0.000161 & 0.69 & 1.388 & 12.44 & 16.85 & 15.52 & --1.64 & --5.94 & --0.63 & 2.20 \\
7 & 0.000148 & 0.75 & 1.183 & 8.53 & 14.60 & 13.55 & --3.93 & --2.81 & --0.59 & 2.44 \\
\hline                
9 & 0.000298 & 0.48 & 1.441 & 16.01 & 15.02 & 14.87 & 0.00 & 0.65 & --0.69 & 1.83 \\
9 & 0.000255 & 0.56 & 1.187 & 8.38 & 9.18 & 8.90 & --0.75 & --0.69 & --0.37 & 1.97 \\
9 & 0.000227 & 0.63 & 1.397 & 6.02 & 5.22 & 4.77 & --0.21 & --0.87 & --0.51 & 1.83 \\
9 & 0.000207 & 0.69 & 1.149 & 5.30 & 8.94 & 8.16 & --1.96 & --2.28 & --0.56 & 1.99 \\
\end{longtable}
\end{center}
\newpage
\begin{center}          
\scriptsize                
\begin{longtable}{|r|c|c||c|c|r|r||r|r|r|r|}             
\hline                 
$Q^2$ & $x$ & $y$ & $\tilde{\sigma}$ & $\delta_{\rm stat}$ &$\delta_{\rm sys}$ & $\delta_{\rm unc}$ & $\delta_{\gamma p}$ &$\delta_{E_{\rm had}}$ & $\delta_{\rm diff}$ & $\delta_{\rm hits}$\\  
(GeV$^2$) & & & MER & (\%) & (\%)  & (\%) & (\%) & (\%) & (\%) & (\%) \\\hline 
5 & 0.000126 & 0.63 & 1.189 & 13.57 & 13.03 & 12.97 & 0.00 & --0.37 & --0.39 & 1.09 \\
\hline                
7 & 0.000278 & 0.40 & 1.123 & 12.50 & 13.23 & 13.19 & 0.00 & --0.63 & --0.05 & 0.74 \\
7 & 0.000231 & 0.48 & 1.173 & 9.63 & 9.46 & 9.40 & --0.11 & --0.38 & 0.13 & 1.00 \\
7 & 0.000198 & 0.56 & 1.229 & 9.74 & 14.02 & 13.93 & --0.93 & 0.54 & --0.21 & 1.12 \\
7 & 0.000176 & 0.63 & 1.225 & 10.51 & 12.08 & 12.01 & --0.16 & --0.65 & --0.35 & 1.03 \\
7 & 0.000161 & 0.69 & 1.650 & 11.15 & 11.36 & 11.17 & --0.38 & 1.67 & --0.36 & 1.07 \\
\end{longtable}
\end{center}

\newpage
\begin{center}          
\scriptsize                
\begin{longtable}{|r|c|c||c|c|r|r||r|r|r|r|}             
\hline                 
$Q^2$ & $x$ & $y$ & $\tilde{\sigma}$ & $\delta_{\rm stat}$ &$\delta_{\rm sys}$ & $\delta_{\rm unc}$ & $\delta_{\gamma p}$ &$\delta_{E_{\rm had}}$ & $\delta_{\rm diff}$ & $\delta_{\rm hits}$\\  
(GeV$^2$) & & & LER & (\%) & (\%)  & (\%) & (\%) & (\%) & (\%) & (\%) \\\hline 
7 & 0.000219 & 0.63 & 1.543 & 14.72 & 12.67 & 12.43 & --0.63 & --1.12 & --0.80 & 1.92 \\
7 & 0.000200 & 0.69 & 1.204 & 9.26 & 9.98 & 9.34 & --1.51 & --2.24 & --0.70 & 2.14 \\
7 & 0.000184 & 0.75 & 1.332 & 5.72 & 8.80 & 7.20 & --2.26 & --3.93 & --0.69 & 2.14 \\
\hline                
9 & 0.000370 & 0.48 & 1.259 & 11.80 & 8.72 & 8.50 & --0.28 & --0.16 & --0.35 & 1.89 \\
9 & 0.000317 & 0.56 & 1.053 & 7.02 & 6.48 & 6.14 & --0.27 & --0.64 & --0.28 & 1.93 \\
9 & 0.000282 & 0.63 & 1.294 & 4.58 & 4.44 & 3.57 & --0.41 & --1.77 & --0.42 & 1.87 \\
\end{longtable}
\end{center}
\newpage
\begin{center}          
\scriptsize                
\begin{longtable}{|r|c|c||c|c|r|r||r|r|r|r|}             
\hline                 
$Q^2$ & $x$ & $y$ & $\tilde{\sigma}$ & $\delta_{\rm stat}$ &$\delta_{\rm sys}$ & $\delta_{\rm unc}$ & $\delta_{\gamma p}$ &$\delta_{E_{\rm had}}$ & $\delta_{\rm diff}$ & $\delta_{\rm hits}$\\  
(GeV$^2$) & & & LER & (\%) & (\%)  & (\%) & (\%) & (\%) & (\%) & (\%) \\\hline 
5 & 0.000176 & 0.56 & 1.022 & 10.84 & 7.82 & 7.74 & --0.28 & --0.54 & --0.37 & 0.90 \\
5 & 0.000157 & 0.63 & 1.011 & 10.04 & 9.90 & 9.76 & --0.27 & --1.09 & --0.34 & 1.10 \\
5 & 0.000143 & 0.69 & 1.049 & 10.23 & 9.43 & 9.23 & --1.02 & --1.05 & --0.54 & 1.18 \\
\hline                
7 & 0.000446 & 0.31 & 0.952 & 15.23 & 10.10 & 10.08 & 0.00 & 0.43 & --0.26 & 0.35 \\
7 & 0.000346 & 0.40 & 0.964 & 9.54 & 6.63 & 6.57 & 0.00 & 0.33 & --0.07 & 0.83 \\
7 & 0.000288 & 0.48 & 0.870 & 7.67 & 6.58 & 6.49 & --0.18 & --0.18 & --0.21 & 1.03 \\
\end{longtable}
\end{center}

%


\newpage

\begin{table}
\renewcommand{\arraystretch}{1.2}
\hspace*{-4cm}
  \centering
  \resizebox{0.68\textwidth}{!}{\begin{minipage}{\textwidth}
        \begin{tabular}{|r|c||c|c|c||r|r|r|r|r|r|r|r|r|r|}
          \hline
$Q^2$ & $x$ & $\tilde{\sigma}$ & $\tilde \delta_{\rm stat}$ &$\tilde \delta_{\rm unc}$& $\tilde \delta_{\rm eID}$ & $\tilde \delta_{\rm dxdy}$ & $\tilde \delta_{e\theta}$ & $\tilde \delta_{E_e}$ & $\tilde \delta_{\rm hdFC}$ & $\tilde \delta_{\rm hdBC}$ & $\tilde \delta_{\rm hdRC}$ & $\tilde \delta_{\rm hdF}$ & $\tilde \delta_{\rm Bhd}$ & $\tilde \delta_{\gamma p}$ \\
(GeV$^2$) & & ZEUS97 & ($\%$) & ($\%$) & ($\%$) & ($\%$) & ($\%$) & ($\%$) & ($\%$) & ($\%$) & ($\%$) & ($\%$) & ($\%$) & ($\%$) \\  
\hline
9 & 0.00025 & 1.214 & 1.60 &  0.90 & 1.64 & 0.56 & 0.00 & 0.59 & --0.08 & 0.21 & 0.39 & 0.24 & 1.61 & 0.20 \\ 
9 & 0.00040 & 1.086 & 1.64 &  0.92 & 1.38 & 0.59 & 0.00 & --0.44 & 0.08 & 0.18 & --0.36 & 0.10 & 0.63 & --0.18 \\ 
\hline
12 & 0.00033 &  1.270 & 2.61  & 0.96 & 2.21 & 1.26& 0.00  & --0.95 & 0.22 & 0.29 & 0.98 & 0.54 & 6.78 & 0.24 \\ 
12 & 0.00041 &  1.186 & 2.30  & 0.87 & 1.61 & 0.59 & 0.00 & 0.22 & --0.16 & --0.06 & 0.13 & 0.26 & 1.87 & --0.01 \\ 
12 & 0.00054 &  1.106 & 2.30  & 0.87 & 1.61 & 0.59& 0.00  & 0.22 & --0.16 & --0.06 & 0.13 & 0.26 & 1.87 & --0.01 \\ 
\hline
17 & 0.00047 &  1.277 & 1.94  & 0.68 & 1.94 & 0.42& 0.00  & --0.49 & 0.11 & 0.26 & 0.39 & 0.22 & 3.49 & 0.09\\
\hline
        \end{tabular}
   \end{minipage}}
\caption{\it 
The reduced cross sections, $\tilde{\sigma}$, for the reaction $e^+p\rightarrow e^+X$, at $\sqrt{s}=300$\gev measured in the years 1996 and 1997, after the adjustments of the binning for the $F_L$ extraction. The first two columns contain the bin centres in $Q^2$ and $x$, the next three contain the measured cross section, the statistical uncertainty and the uncorrelated systematic uncertainty, respectively. The final ten columns list the bin-to-bin correlated uncertainties from each systematic source. The normalisation uncertainty (see Section \ref{sec:9697}) is not included.}
\label{tab:sig9697}
\end{table}


\begin{table}
\renewcommand{\arraystretch}{1.5}
\scriptsize
\begin{center}
\begin{tabular}{|r|c||c|c||c|c|}
\hline
$Q^{2}$ & $x$& $F_{2}$ & $F_{2}$ & $F_{L}$  & $F_{L}$ \\
 (GeV$^2$) &  &  (unconstrained fit) &  (constrained fit) & (unconstrained fit) & (constrained fit) \\
\hline
9 & 0.00025 & 1.223$^{+0.045}_{-0.039}$ & 1.259$^{+0.039}_{-0.033}$ & --0.06$^{+0.22}_{-0.19}$ & 0.00$^{+0.24}_{-0.00}$ \\
9 & 0.00031 & 1.325$^{+0.111}_{-0.129}$ & 1.331$^{+0.111}_{-0.081}$ & 0.25$^{+0.55}_{-0.42}$ & 0.47$^{+0.29}_{-0.41}$ \\  
9 & 0.00040 & 1.085$^{+0.039}_{-0.033}$ & 1.115$^{+0.033}_{-0.027}$ & --0.08$^{+0.53}_{-0.51}$ & 0.00$^{+0.57}_{-0.00}$ \\
\hline
12 & 0.00033 & 1.319$^{+0.063}_{-0.057}$ &  1.349$^{+0.063}_{-0.051}$ & 0.15$^{+0.20}_{-0.20}$ & 0.29$^{+0.20}_{-0.16}$ \\
12 & 0.00041 & 1.181$^{+0.045}_{-0.039}$ & 1.229$^{+0.039}_{-0.027}$ & --0.24$^{+0.26}_{-0.20}$ & 0.00$^{+0.17}_{-0.00}$ \\ 
12 & 0.00054 & 1.139$^{+0.045}_{-0.039}$ &1.169$^{+0.033}_{-0.033}$ & --0.09$^{+0.44}_{-0.40}$ & 0.00$^{+0.46}_{-0.00}$ \\  
\hline
17 & 0.00047 & 1.343$^{+0.039}_{-0.039}$ & 1.361$^{+0.045}_{-0.033}$ & 0.18$^{+0.17}_{-0.11}$ & 0.31$^{+0.12}_{-0.14}$ \\
17 & 0.00058 & 1.217$^{+0.057}_{-0.045}$ & 1.271$^{+0.039}_{-0.027}$ & --0.20$^{+0.22}_{-0.20}$ &0.00$^{+0.15}_{-0.00}$ \\  
17 & 0.00076 & 1.175$^{+0.033}_{-0.045}$ & 1.199$^{+0.033}_{-0.027}$ & --0.09$^{+0.29}_{-0.35}$ &0.00$^{+0.30}_{-0.00}$ \\  
\hline
24 & 0.00067 & 1.373$^{+0.033}_{-0.027}$ & 1.385$^{+0.033}_{-0.027}$ & 0.21$^{+0.13}_{-0.11}$ & 0.29$^{+0.12}_{-0.10}$ \\
24 & 0.00082 & 1.277$^{+0.033}_{-0.027}$ & 1.289$^{+0.027}_{-0.027}$ & 0.06$^{+0.16}_{-0.13}$ & 0.14$^{+0.09}_{-0.12}$ \\ 
24 & 0.00108 & 1.193$^{+0.033}_{-0.027}$ & 1.205$^{+0.033}_{-0.021}$ & 0.09$^{+0.28}_{-0.22}$ & 0.19$^{+0.19}_{-0.17}$ \\ 
\hline
32 & 0.00089 & 1.355$^{+0.027}_{-0.027}$ & 1.361$^{+0.033}_{-0.021}$ & 0.06$^{+0.11}_{-0.11}$ &  0.14$^{+0.08}_{-0.11}$ \\  
32 & 0.00109 & 1.271$^{+0.027}_{-0.021}$ & 1.283$^{+0.027}_{-0.021}$ & 0.13$^{+0.16}_{-0.11}$ & 0.22$^{+0.12}_{-0.12}$ \\  
32 & 0.00143 & 1.193$^{+0.027}_{-0.027}$ & 1.205$^{+0.021}_{-0.021}$ & --0.02$^{+0.23}_{-0.20}$ &0.00$^{+0.23}_{+0.00}$ \\
\hline
45 & 0.00125 & 1.319$^{+0.027}_{-0.027}$ & 1.325$^{+0.027}_{-0.021}$ & 0.12$^{+0.13}_{-0.10}$ & 0.19$^{+0.12}_{-0.09}$ \\ 
45 & 0.00153 & 1.259$^{+0.027}_{-0.027}$ & 1.271$^{+0.027}_{-0.021}$ & 0.06$^{+0.14}_{-0.16}$ & 0.09$^{+0.11}_{-0.09}$ \\ 
45 & 0.00202 & 1.163$^{+0.027}_{-0.021}$ & 1.169$^{+0.027}_{-0.021}$ & 0.30$^{+0.20}_{-0.23}$ & 0.35$^{+0.17}_{-0.21}$ \\ 
\hline
60 & 0.00167 & 1.319$^{+0.033}_{-0.021}$ & 1.331$^{+0.027}_{-0.027}$ & 0.12$^{+0.14}_{-0.11}$ & 0.19$^{+0.12}_{-0.11}$ \\ 
60 & 0.00204 & 1.235$^{+0.027}_{-0.021}$ & 1.241$^{+0.027}_{-0.021}$ & 0.22$^{+0.16}_{-0.17}$ & 0.25$^{+0.16}_{-0.14}$ \\
60 & 0.00269 & 1.133$^{+0.027}_{-0.021}$ & 1.145$^{+0.027}_{-0.015}$ & 0.04$^{+0.25}_{-0.23}$ & 0.09$^{+0.21}_{-0.09}$ \\ 
\hline
80 & 0.00222 & 1.235$^{+0.033}_{-0.027}$ & 1.247$^{+0.027}_{-0.027}$ & 0.15$^{+0.13}_{-0.14}$ & 0.17$^{+0.11}_{-0.11}$ \\ 
80 & 0.00272 & 1.223$^{+0.033}_{-0.021}$ & 1.229$^{+0.033}_{-0.021}$ & 0.45$^{+0.14}_{-0.20}$ & 0.44$^{+0.19}_{-0.16}$ \\
80 & 0.00359 & 1.097$^{+0.027}_{-0.021}$ & 1.103$^{+0.027}_{-0.015}$  & 0.18$^{+0.25}_{-0.23}$ & 0.23$^{+0.20}_{-0.18}$ \\
\hline
110 & 0.00306 & 1.187$^{+0.033}_{-0.033}$ & 1.199$^{+0.033}_{-0.027}$ & 0.12$^{+0.16}_{-0.17}$ &0.16$^{+0.12}_{-0.14}$ \\  
110 & 0.00374 & 1.157$^{+0.033}_{-0.033}$ & 1.163$^{+0.033}_{-0.027}$ & 0.21$^{+0.20}_{-0.22}$ &0.20$^{+0.20}_{-0.14}$ \\  
110 & 0.00493 & 1.037$^{+0.027}_{-0.021}$ & 1.049$^{+0.027}_{-0.015}$ & --0.12$^{+0.31}_{-0.23}$ &0.00$^{+0.25}_{+0.00}$ \\  
\hline

\end{tabular}
\caption{\it Extracted values of $F_2$ and $F_L$ at 27 $(x,Q^2)$ points. Values are quoted for the unconstrained and constrained fits. 
The quoted uncertainties cover both the statistical and systematic sources. The normalisation uncertainty (see Section \ref{sec:fits}) is not included. For the constrained fits with mode at $F_L=0$, the upper uncertainties correspond to 68$\%$ probability limits.
}

\label{tab:F2FLxQ2}
\end{center}
\end{table}


\begin{table}
\renewcommand{\arraystretch}{1.7}
\scriptsize
\begin{center}
\begin{tabular}{|r||c|c||c|c|}
\hline
$Q^{2}$  & $F_{L}(Q^2)$ & $F_{L}(Q^2)$ & $R(Q^2)$ & $R(Q^2)$ \\
 (GeV$^2$) & (unconstrained fit) & (constrained fit) & (unconstrained fit) & (constrained fit)\\
\hline
 9 & --0.040$_{-0.167}^{+0.191}$ & 0.000$_{-0.000}^{+0.135}$ & 0.01$_{-0.13}^{+0.17}$ & 0.05$_{-0.05}^{+0.17}$ \\
\hline
12 & --0.003$_{-0.162}^{+0.181}$ & 0.000$_{-0.000}^{+0.135}$ & 0.01$_{-0.12}^{+0.17}$ & 0.05$_{-0.05}^{+0.15}$\\
\hline
17 & 0.075$_{-0.104}^{+0.130}$ & 0.126$_{-0.102}^{+0.090}$ & 0.09$_{-0.10}^{+0.11}$& 0.14$_{-0.11}^{+0.08}$ \\
\hline
24 & 0.160$_{-0.115}^{+0.095}$ & 0.184$_{-0.078}^{+0.107}$ & 0.14$_{-0.07}^{+0.11}$& 0.17$_{-0.08}^{+0.10}$ \\
\hline
32 & 0.078$_{-0.093}^{+0.093}$ & 0.102$_{-0.076}^{+0.077}$ & 0.08$_{-0.07}^{+0.08}$ & 0.09$_{-0.05}^{+0.09}$\\
\hline
45 & 0.115$_{-0.090}^{+0.111}$ & 0.138$_{-0.075}^{+0.102}$  & 0.10$_{-0.06}^{+0.10}$& 0.12$_{-0.08}^{+0.09}$ \\ 
\hline
60 & 0.135$_{-0.091}^{+0.132}$ & 0.179$_{-0.102}^{+0.090}$ & 0.14$_{-0.09}^{+0.10}$ & 0.16$_{-0.09}^{+0.10}$\\
\hline
80 & 0.222$_{-0.108}^{+0.127}$ & 0.244$_{-0.099}^{+0.112}$ & 0.25$_{-0.12}^{+0.12}$& 0.25$_{-0.11}^{+0.14}$ \\
\hline
110 & 0.086$_{-0.100}^{+0.154}$ &0.101$_{-0.079}^{+0.127}$ & 0.10$_{-0.10}^{+0.16}$ & 0.12$_{-0.09}^{+0.13}$ \\ 
\hline 
\end{tabular}
\caption{\it Extracted values of $F_L$ and $R$ at 9 $Q^2$ points. Values are quoted for the unconstrained and constrained fits. 
The quoted uncertainties cover both the statistical and systematic sources. The normalisation uncertainty (see Section \ref{sec:fits}) is not included. For the constrained fits with mode at $F_L=0$, the upper uncertainties correspond to 68$\%$ probability limits.}

\label{tab:FLRQ2}
\end{center}
\end{table}


\newpage


\end{document}